\documentclass[
 reprint,
 amsmath,amssymb,
 aps,
 prl,
]{revtex4-2}

\usepackage{graphicx}
\usepackage{dcolumn}
\usepackage{bm}
\usepackage{orcidlink}

\begin{document}

\title{Generalized Fourier's law in mesoscopic systems}

\author{Luciano C. Lapas\orcidlink{0000-0002-0029-5625}}
\email{luciano.lapas@unila.edu.br}
\affiliation{Latin American Institute of Life and Natural Sciences, Universidade Federal da Integra\c{c}\~ao Latino-Americana, Av. Tancredo Neves, 6731, Bloco 6, Espa\c{c}o 3, Sala 5, Foz do Igua\c{c}u, 85867-970, PR, Brazil.}

\date{\today}

\begin{abstract}
Fourier's law fails when the mean free path of the energy carriers becomes comparable to the length and time scales over which the temperature field varies. We derive a thermodynamically consistent generalization in which the conductivity becomes a nonlocal memory operator $\bm{\kappa}_{\mathrm{eff}}(\mathbf{k},\omega)$, by combining mesoscopic nonequilibrium thermodynamics with the Mori--Kubo--Zwanzig projection-operator formalism. The Onsager kernel is a tensor sum over vibrational normal modes weighted by their Bose heat capacities and relaxation functions; it satisfies the second law for any closure consistent with the fluctuation--dissipation theorem and reduces, in the relaxation-time closure, to the nonlocal conductivity of the phonon Boltzmann equation. Two consequences follow. First, direction-dependent sampling of this tensor in a pump--probe experiment makes a nominally isotropic crystal exhibit apparent conductivities $\Lambda_z\neq\Lambda_r$. Second, the modulation frequency does not introduce temporal memory but sets the probed wavevector through the thermal penetration depth, so that the frequency-dependent suppression of the apparent conductivity measured by time-domain thermoreflectance on Ge and SiGe alloys is a spatial-nonlocality effect, with effective lengths of $0.25$--$0.4~\mu$m. With published mean-free-path spectra and no parameter adjusted to the thermoreflectance data, the kernel reproduces the spot-size dependence of Si, the frequency dependence of the apparent conductivity of Ge and Si$_{0.2}$Ge$_{0.8}$ and the rise of the apparent interface conductance of Al/Si$_{0.99}$Ge$_{0.01}$, and contains the two-channel ballistic/diffusive model as a two-sector projection.
\end{abstract}

\maketitle

\textit{Introduction.---}The classical theory of heat conduction rests on Fourier's law,
\begin{equation}
 \mathbf{J}_q = -\kappa \nabla T,
 \label{eq:FourierLaw}
\end{equation}
relating the heat flux $\mathbf{J}_q$ to the local temperature gradient through a scalar conductivity $\kappa$. Equation~\eqref{eq:FourierLaw} is accurate whenever the carrier mean free path (MFP) $\ell_c$ is far shorter than the experimental length scale $L_{\mathrm{exp}}$, i.e.\ in the diffusive regime where the Knudsen number $\mathrm{Kn}\equiv \ell_c/L_{\mathrm{exp}}\ll 1$ and carriers (e.g., phonons in dielectrics and most semiconductors, electrons in metals~\cite{Chen2005,Greffet2007b}) relax to local equilibrium much faster than the macroscopic field evolves. Local energy conservation then yields the diffusion equation $\partial_t T = a\,\nabla^2 T$~\cite{Fourier1822,Greffet2007a}, with $a$ the thermal diffusivity, and a localized pulse spreads into a Gaussian profile whose width grows as $2\sqrt{at}$. This description breaks down when $L_{\mathrm{exp}}$ is no longer large compared with the longest active MFP, as in nanostructures~\cite{Bimberg1999}, colloids~\cite{Berne2000}, and biological organelles~\cite{Alberts2022}, where $\mathrm{Kn}\gtrsim\mathcal{O}(1)$ marks the onset of nonlocal, partially ballistic transport~\cite{Cahill2003,Greffet2007b}. Quantitative evidence is now abundant: quasiballistic transport and size-dependent effective conductivity in silicon membranes~\cite{Johnson2013}; the anisotropic failure of Fourier's law in time-domain thermoreflectance (TDTR) on Si and SiGe~\cite{Wilson2014}; and non-Gaussian temperature profiles in translucent polymers and glasses, where internal infrared radiation provides an additional transport channel with millimetre-scale effective MFPs~\cite{Zheng2024}.

These breakdowns are not all of the same kind, and a useful theory must distinguish them. The general constitutive response is nonlocal in both time and space, but a given experiment probes one axis preferentially. In TDTR the relevant timescale $1/f$ is far longer than carrier lifetimes, so temporal memory is negligible and the modulation frequency $f$ acts only by fixing a spatial scale, the thermal penetration depth $d_p=\sqrt{\kappa_{\mathrm{ref}}/\pi C f}$, with $C$ the volumetric heat capacity~\cite{Wilson2014} and $\kappa_{\mathrm{ref}}$ a fixed reference conductivity (the bulk value below); the same holds for the macroscale radiative anomaly of Ref.~\cite{Zheng2024}. Temporal memory dominates instead in second-sound experiments, whereas room-temperature transient gratings, like TDTR, probe spatial nonlocality~\cite{Maznev2011}. The TDTR breakdown is moreover \emph{anisotropic}: beam-offset measurements on nominally isotropic Si return distinct through-plane and in-plane conductivities, $\Lambda_z\approx 140$ and $\Lambda_r\approx 80~\mathrm{W\,m^{-1}\,K^{-1}}$~\cite{Wilson2014,Feser2012}, and it is interface mediated, requiring a nonequilibrium resistance beyond a local boundary conductance~\cite{Wilson2014}. Following Wilson and Cahill, we use $\Lambda_A$, $\Lambda_z$, and $\Lambda_r$ to denote \emph{apparent} conductivities extracted by fitting TDTR data with local diffusive models, while $\kappa_{\mathrm{bulk}}$ denotes the intrinsic bulk conductivity in the Fourier limit. Related nonlocal approaches include phonon Boltzmann transport equation (BTE) kernels in the relaxation-time approximation~\cite{Mahan1988,Koh2014,Hua2014,Allen2018} and beyond it~\cite{Hua2020}, applied to TDTR analytically~\cite{Koh2014,YangDames2015}, through mean-free-path spectroscopy~\cite{Minnich2011}, by Monte Carlo~\cite{Ding2014}, or with measured interfacial transmissivities~\cite{Hua2019}; Guyer--Krumhansl hydrodynamic analyses~\cite{Beardo2022,Xiang2022}; two-channel interface models~\cite{Wilson2013}; and, closest to the present work, a spatiotemporal Green--Kubo kernel obtained by Zwanzig projection and realized for phonons through the linearized BTE~\cite{Zeng2026}. The kinetic treatments start from a prescribed collision operator, so that their second law and exponential memory are those of that operator, and the interface enters them through a boundary condition, fitted rates, or a separable kernel with illustrative parameters. The constitutive law derived here starts instead from mesoscopic entropy production and the Mori--Kubo--Zwanzig identity: it guarantees nonnegative entropy production for any closure consistent with the fluctuation--dissipation theorem, recovers the BTE and Guyer--Krumhansl kernels as Markovian and single-pole limits of a memory function that obeys this theorem, and, given a modal boundary condition, yields the interfacial nonequilibrium resistance from the same kernel, the two-channel model~\cite{Wilson2013} being its two-sector projection [Supplemental Material (SM)~\cite{SM}, Sec.~SIX].

In this Letter we derive this constitutive law. Combining mesoscopic nonequilibrium thermodynamics (MNET)~\cite{deGroot1984,Vilar2001,Reguera2005} with the Mori--Kubo--Zwanzig projection-operator formalism~\cite{Mori1965,Kubo1957,Zwanzig1961,Zwanzig2001} promotes the conductivity to a nonlocal memory operator $\bm{\kappa}_{\mathrm{eff}}(\mathbf{k},\omega)$ whose Onsager kernel decomposes into vibrational normal modes weighted by Bose heat capacities. We show that this single kernel (i) produces apparent anisotropy $\Lambda_z\neq\Lambda_r$ as a consequence of the directional modal weights, (ii) accounts for the TDTR conductivities of Si, Ge and Si$_{1-x}$Ge$_x$ as a spatial-nonlocality effect set by $d_p(f)$: in its parameter-free relaxation-time closure, with first-principles-anchored MFP spectra, it reproduces the spot-size dependence and in-plane anisotropy of Si and the frequency dependence of Ge and Si$_{0.2}$Ge$_{0.8}$ (SM, Secs.~SIV.B--SVIII), and (iii) contains the two-channel ballistic/diffusive description of nondiffusive transport~\cite{Wilson2014,Armstrong1981,Maznev2011} as a two-sector projection.

\textit{Generalized constitutive relation.---}We extend the validity of the Gibbs equation to internal degrees of freedom $\bm{\Gamma}=(\mathbf{p},\mathbf{r})$ that have not yet relaxed to local equilibrium. For a phase-space density $\rho(\bm{\Gamma},t)$ of heat carriers, the entropy production rate is $\sigma=-\int \mathbf{J}_{\bm{\Gamma}}\!\cdot\!\nabla_{\bm{\Gamma}}(\mu/T)\,d\bm{\Gamma}\ge 0$, with $\mu(\bm{\Gamma},t)$ a nonequilibrium chemical potential~\cite{Vilar2001,Reguera2005}. Even in steady state, bosonic carriers can acquire a nonzero effective chemical potential, as in near-field photonic cooling~\cite{Zhu2019} and nonequilibrium Casimir experiments~\cite{Chen2016}.

Postulating a causal linear response in phase space consistent with $\sigma\ge0$, and coarse-graining onto the energy-density field, yields a generalized constitutive law for the macroscopic heat flux (see SM~\cite{SM}),
\begin{equation}
\mathbf{J}_q(\mathbf{r}, t) = \!\int_{-\infty}^{t}\!\! dt'\! \int d\mathbf{r}' \, \mathbf{L}_{qq}(\mathbf{r},\mathbf{r}',t-t') \,\nabla_{\mathbf{r}'}\!\!\left[\frac{1}{T(\mathbf{r}', t')}\right].
\label{eq:nonlocal-heat-flux}
\end{equation}
The spatiotemporal kernel $\mathbf{L}_{qq}$ is not {\it ad hoc}: it arises from the projection of phase-space correlations connecting transport at $\mathbf{r}'$ to $\mathbf{r}$ through carrier propagation and scattering, and its symmetric part is positive semidefinite, guaranteeing $\sigma\ge0$~\cite{SM}. Linearizing about a reference $T_0$ via $\nabla(1/T)\simeq -T_0^{-2}\nabla\delta T$ and Fourier transforming in space and time recasts Eq.~\eqref{eq:nonlocal-heat-flux} as a generalized Fourier law,
\begin{equation}
\mathbf{J}_q(\mathbf{k},\omega) = - i\,\bm{\kappa}_{\mathrm{eff}}(\mathbf{k},\omega)\cdot\mathbf{k}\,\, \delta T(\mathbf{k},\omega),
\label{eq:kappa-k-omega}
\end{equation}
with $\bm{\kappa}_{\mathrm{eff}}(\mathbf{k},\omega)\equiv \widetilde{\mathbf{L}}_{qq}(\mathbf{k},\omega)/T_0^2$ the generalized conductivity tensor. The classical law, Eq.~\eqref{eq:FourierLaw}, is recovered strictly in the hydrodynamic limit $\mathbf{k}\!\to\!0,\ \omega\!\to\!0$, where the kernel reduces to a Dirac delta in space and time and $\bm{\kappa}_{\mathrm{eff}}\!\to\!\kappa\,\mathbf{1}$. In TDTR the surface response is a Hankel-space integral over in-plane wavevectors weighted by the beam profiles~\cite{Feser2012,Cahill2004}, so that the in-plane kernel enters that integral, $\kappa_r\to\kappa_{\mathrm{eff}}^{rr}(k)$, without a single-wavevector approximation (to leading order in $k_z/k_r$, away from the transducer interface); the dominant gradients are nevertheless set by $k_z\sim d_p^{-1}(f)$ through-plane and $k_r\sim w_0^{-1}$ in-plane, making the fitted local conductivities directional proxies for $\bm{\kappa}_{\mathrm{eff}}(\mathbf{k})$~\cite{Wilson2014}.

\textit{Modal kernel and anisotropy.---}To ground the kernel microscopically we project the Liouville evolution of $\bm{\Gamma}$ onto a column vector $\mathbf{A}(t)$ of slow variables. The Mori--Kubo--Zwanzig identity yields the exact generalized Langevin equation (GLE)~\cite{SM},
\begin{equation}
\frac{d\mathbf{A}(t)}{dt}= i\bm{\Omega}\,\mathbf{A}(t) - \!\int_0^t\! d\tau\, \bm{\Phi}(\tau)\,\mathbf{A}(t-\tau) + \mathbf{F}(t),
\label{eq:GLE-main}
\end{equation}
with $\bm{\Omega}$ the reversible frequency matrix, $\mathbf{F}(t)$ the projected stochastic force, and the memory kernel obeying the fluctuation--dissipation theorem $\bm{\Phi}(\tau)=\langle\mathbf{F}(\tau)\mathbf{F}^{T}(0)\rangle\,\mathbf{C}^{-1}$, $\mathbf{C}\equiv\langle\mathbf{A}\mathbf{A}^{T}\rangle$. The normalized propagator $\mathbf{R}(t)=\mathbf{C}(t)\mathbf{C}^{-1}$ has Laplace transform $\widetilde{\mathbf{R}}(z)=[z\mathbf{1}-i\bm{\Omega}+\widetilde{\bm{\Phi}}(z)]^{-1}$, whose poles set the mesoscopic timescales.

\begin{figure}[t]
 \centering
 \includegraphics[width=0.95\linewidth]{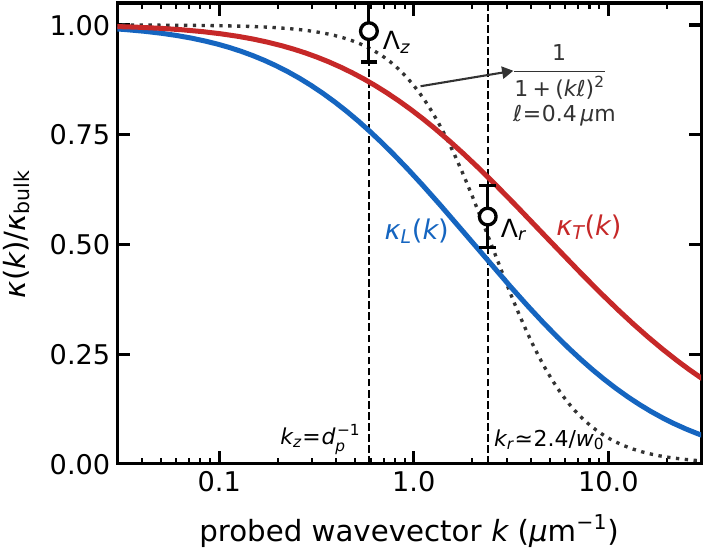}
 \caption{Apparent anisotropy from a single nonlocal kernel. Longitudinal and transverse blocks $\kappa_L(k)$, $\kappa_T(k)$ of the relaxation-time kernel of Si at 300~K, Eqs.~\eqref{eq:Lqq-modal}--\eqref{eq:kappa-tensor} with a first-principles-anchored MFP spectrum (SM, Secs.~SIV.B and SVIII; no adjustable parameter), and the one-length Lorentzian of Eq.~\eqref{eq:kspace} with $\ell=0.4~\mu$m (dotted). A TDTR experiment samples the through-plane direction at $k_z\simeq d_p^{-1}(f)$ and the in-plane direction at $k_r\simeq2.4/w_0$ (dashed lines; $f=9.8$~MHz, $w_0=1~\mu$m), so a nominally isotropic crystal yields $\Lambda_z\neq\Lambda_r$; symbols are the beam-offset values of Wilson and Cahill~\cite{Wilson2014}, normalized to the measured $\kappa_{\mathrm{bulk}}=142~\mathrm{W\,m^{-1}\,K^{-1}}$ (the kernel to its own bulk value, 148). The through-plane point lies above $\kappa_L(k_z)$ because the Si response in that direction is interface-dominated (SM, Sec.~SIV.C).}
 \label{fig:concept}
\end{figure}

Choosing the components of $\mathbf{A}$ as the modal energy fluctuations $A_\mu(t)=\delta E_\mu(t)=\varepsilon_\mu[n_\mu(t)-n_\mu^{\mathrm{eq}}]$ of the vibrational eigenmodes $\phi_\mu(\mathbf{r})$ of the confined medium (with $\varepsilon_\mu=\hbar\omega_\mu$), the heat flux separates into geometry and dynamics, $\mathbf{J}_q(\mathbf{r},t)=\sum_\mu \bm{\Psi}_\mu(\mathbf{r})\,A_\mu(t)$, with the \emph{vectorial} mode flux functions
\begin{equation}
\bm{\Psi}_\mu(\mathbf{r})=\mathbf{v}_\mu(\mathbf{r})\,|\phi_\mu(\mathbf{r})|^2
\label{eq:Psi}
\end{equation}
carrying the directional group velocity $\mathbf{v}_\mu$. The Onsager kernel then follows in closed, tensor form~\cite{SM},
\begin{equation}
L_{qq}^{ij}(\mathbf{r},\mathbf{r}',t)=T_0^{2}\sum_{\mu,\nu}\Psi_\mu^{i}(\mathbf{r})\,\Psi_\nu^{j}(\mathbf{r}')\,R_{\mu\nu}(t)\,c_{\mathrm{ph}}(\omega_\nu,T),
\label{eq:Lqq-modal}
\end{equation}
where the classical variance is replaced by the spectral (Bose) heat capacity
\begin{equation}
c_{\mathrm{ph}}(\omega,T)=k_B\!\left(\frac{\hbar\omega}{2k_BT}\right)^{2}\!\mathrm{csch}^2\!\left(\frac{\hbar\omega}{2k_BT}\right).
\label{eq:cph}
\end{equation}
Equation~\eqref{eq:Lqq-modal} is the central structural result. In a confined structure spatial nonlocality is carried by the mode overlap $\Psi_\mu^{i}(\mathbf{r})\Psi_\nu^{j}(\mathbf{r}')$; in a bulk crystal, where $|\phi_\mu|^2$ is uniform, it is carried by the reversible streaming term $i\Omega_{\mu\nu}(\mathbf{k})=-i\,\mathbf{k}\cdot\mathbf{v}_\mu\delta_{\mu\nu}$ of Eq.~\eqref{eq:GLE-main}, so that $R_{\mu\nu}$ acquires a $\mathbf{k}$ dependence; temporal memory resides in the dissipative kernel $\bm{\Phi}$; and the quantum weight $c_{\mathrm{ph}}(\omega_\nu,T)$ ``freezes out'' high-frequency modes for $k_BT\ll\hbar\omega_\nu$, correcting the systematic overestimate of heat flow by classical models below the Debye temperature~\cite{Schwab2000,Cahill2003}, and coincides with the modal heat capacity $C_\mu$ of kinetic theory~\cite{Hua2019,Zeng2026}. With the Markovian closure $\Phi_{\mu\nu}=2\tau_\mu^{-1}\delta_{\mu\nu}\delta(t)$, Eqs.~\eqref{eq:GLE-main} and \eqref{eq:Lqq-modal} give $\kappa^{ij}_{\mathrm{eff}}(\mathbf{k})=V^{-1}\sum_\mu C_\mu v^i_\mu v^j_\mu\tau_\mu/[1+(\mathbf{k}\cdot\bm{\Lambda}_\mu)^2]$, with $\bm{\Lambda}_\mu=\mathbf{v}_\mu\tau_\mu$ and $V$ the volume, which is the nonlocal conductivity of the Boltzmann equation~\cite{Mahan1988,Hua2019}; its longitudinal and transverse blocks are suppressed as $1-\tfrac35(k\Lambda)^2$ and $1-\tfrac15(k\Lambda)^2$ at long wavelength (Fig.~\ref{fig:concept}; SM, Sec.~SIV.B; scope and limitations in SM, Sec.~SVIII.A). Distinct carrier families (acoustic and optical phonons, or internal radiation in translucent solids~\cite{Zheng2024}) enter additively through disjoint mode sectors.

The tensor structure of Eq.~\eqref{eq:Lqq-modal} has an immediate, testable consequence. Fourier transforming in space and time, the diagonal blocks of the conductivity read
\begin{equation}
\kappa_{\mathrm{eff}}^{ii}(\mathbf{k},\omega)=\frac{1}{V}\sum_{\mu,\nu}\widetilde\Psi_\mu^{i}(\mathbf{k})\,\widetilde\Psi_\nu^{i}(-\mathbf{k})\,\widetilde R_{\mu\nu}(\mathbf{k},\omega)\,c_{\mathrm{ph}}(\omega_\nu,T),
\label{eq:kappa-tensor}
\end{equation}
where the mode overlap (confined modes) or the streaming propagator $\widetilde R_{\mu\nu}(\mathbf{k},\omega)$ (bulk) is maximal at $\mathbf{k}\!\to\!0$ and is cut off for $|\mathbf{k}|\gtrsim\ell^{-1}$, so each block decreases monotonically with $|\mathbf{k}|$. Because $\bm{\Psi}_\mu$ points along the mode group velocity, the kernel is a tensor whose blocks a TDTR experiment samples at direction-dependent wavevectors, $k_z\!\simeq d_p^{-1}(f)=\sqrt{\pi C f/\kappa_{\mathrm{ref}}}$ and $k_r\!\simeq\! c/w_0$ ($c\simeq2.3$--$2.4$ for the beam-offset width at $w_0\approx1$--$1.5~\mu$m, rising slowly with $w_0$; SM, Sec.~SVIII); this sampling yields, in the absence of interfacial effects,
\begin{equation}
\Lambda_z=\kappa_{\mathrm{eff}}^{zz}\!\big(k_z\!\sim\! d_p^{-1}\big)\neq \Lambda_r=\kappa_{\mathrm{eff}}^{rr}\!\big(k_r\!\sim\! c/w_0\big),
\label{eq:anisotropy}
\end{equation}
with the more strongly sampled direction the more suppressed (Fig.~\ref{fig:concept}). For Si at $w_0\approx1\,\mu$m and $f=9.8$~MHz, with $\kappa_{\mathrm{ref}}\approx142$~$\mathrm{W\,m^{-1}\,K^{-1}}$ (taken as the bulk value) and $C\approx1.6\times10^{6}~\mathrm{J\,m^{-3}\,K^{-1}}$, one finds $d_p\approx1.7\,\mu$m, hence $k_z\approx0.6<k_r\approx2.4~\mu\mathrm{m}^{-1}$ and $\Lambda_r<\Lambda_z$, in qualitative agreement with the beam-offset values quoted above: a nominally isotropic kernel produces apparent anisotropy purely through direction-dependent sampling. Raising $f$ shifts $k_z$ outward and suppresses the through-plane channel; shrinking $w_0$ shifts $k_r$ outward and suppresses the in-plane channel, reproducing the experimentally observed selectivity of the two knobs~\cite{Wilson2014}. Inserting this parameter-free relaxation-time kernel, built from a first-principles-anchored MFP spectrum, into the multilayer TDTR model~\cite{Cahill2004,Feser2012} reproduces both the pump-diameter dependence $\Lambda_A(D)$ of Ding \emph{et al.}~\cite{Ding2014} ($121\to148$ vs.\ $115\to143~\mathrm{W\,m^{-1}\,K^{-1}}$ for $D=3.7\to60~\mu$m) and the beam-offset $\Lambda_r$ at $w_0=1~\mu$m ($70$--$92$, depending on the MFP spectrum, vs.\ $80~\mathrm{W\,m^{-1}\,K^{-1}}$) [SM, Fig.~S1(b,c) and Sec.~SVIII], whereas through-plane transport in Si is governed by the interfacial condition (below).

\textit{Limiting cases.---}For a bath spectral density $S(\omega)\propto\omega^{s}$, $\widetilde{\bm{\Phi}}(z)\sim z^{s-1}$ classifies normal ($s=1$), anomalous ($0<s<2$, $s\neq1$) and ballistic ($s=2$) transport (SM, Sec.~SV~\cite{SM,Morgado2002,Lapas2007,Vainstein2006,Lapas2008,Flekkoy2021}). A Markovian kernel $\bm{\Phi}(t)\approx2\tau_q^{-1}\mathbf{1}\,\delta(t)$ makes the flux correlations exponential, i.e.\ gives $\widetilde{\mathbf{R}}(z)$ a single inertial pole at $z=-\tau_q^{-1}$: for $\omega\tau_q\ll1$ it recovers Fourier's law (and, with streaming, the Boltzmann kernel above), while retaining the pole yields the Cattaneo--Vernotte equation~\cite{Cattaneo1948,Vernotte1958} $\tau_q\partial_t\mathbf{J}_q+\mathbf{J}_q=-\kappa\nabla T$, whose spatial counterpart is the Guyer--Krumhansl kernel~\cite{Guyer1966,Zeng2026}. Temporal memory ($\omega\tau_q\sim1$) thus governs second sound, whereas TDTR and room-temperature transient gratings~\cite{Maznev2011} operate on timescales far longer than $\tau_q$.

\textit{Spatial nonlocality and the TDTR data.---}To lowest order in gradients the isotropic nonlocal conductivity is represented by a minimal one-length ansatz, a Lorentzian in wavevector,
\begin{equation}
\kappa_{\mathrm{eff}}(k)=\frac{\kappa_{\mathrm{bulk}}}{1+(k\ell)^2},
\label{eq:kspace}
\end{equation}
which is the $[0/1]$ Pad\'e closure consistent with the long-wavelength expansion of a finite-range spatial kernel (SM, Sec.~SIV.A); in the relaxation-time closure $\ell^2=\tfrac35\langle\Lambda^2\rangle_\kappa$, with $\langle\Lambda^2\rangle_\kappa$ the $\kappa$-weighted second moment of the MFP spectrum, but for broad spectra the full kernel decays almost logarithmically in $k$ (Fig.~\ref{fig:concept}), so that Eq.~\eqref{eq:kspace} is an effective one-length description over the measured decade.

Evaluated at $k_z=d_p^{-1}$ this gives a frequency-resolved apparent conductivity
\begin{equation}
\Lambda_A(f)=\frac{\kappa_{\mathrm{bulk}}}{1+f/f^\ast},\qquad f^\ast=\frac{\kappa_{\mathrm{ref}}}{\pi C\,\ell^{2}},
\label{eq:LambdaA}
\end{equation}
controlled by a single effective length $\ell$ rather than a relaxation time. The crossover frequency $f^\ast$ is large (response flat) when $\ell$ is short, and small (early roll-off) when $\ell$ is long.

\begin{figure}[t]
 \centering
 \includegraphics[width=0.95\linewidth]{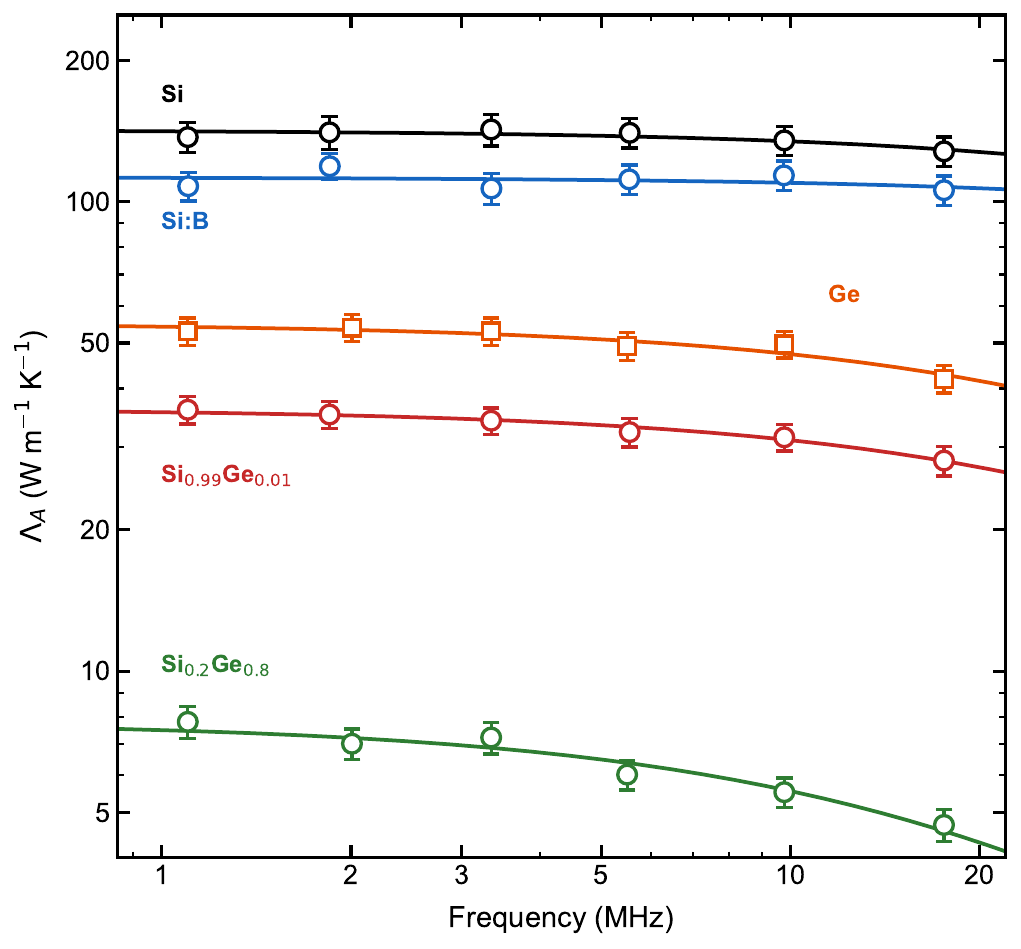}
 \caption{Apparent thermal conductivity $\Lambda_A$ versus modulation frequency: TDTR data of Wilson and Cahill~\cite{Wilson2014} (symbols, with error bars; digitized from their Fig.~3a at $w_0=10~\mu$m) and two-parameter fits of Eq.~\eqref{eq:LambdaA} (solid lines). Fourier's law predicts a constant $\Lambda_A$; Si and Si:B stay flat while Ge and the SiGe alloys are suppressed at high $f$, where the shrinking penetration depth $d_p(f)$ probes a larger wavevector. The effective lengths are $\ell=0.41$, $0.34$ and $0.25~\mu$m for Ge, Si$_{0.99}$Ge$_{0.01}$ and Si$_{0.2}$Ge$_{0.8}$, respectively. Parameter-free kernel calculation of the same data and of the apparent interface conductance $G_A(f)$: SM, Fig.~S2.}
 \label{fig:gle_fit}
\end{figure}

Figure~\ref{fig:gle_fit} compares Eq.~\eqref{eq:LambdaA} with the apparent conductivities of Wilson and Cahill~\cite{Wilson2014} over $f=1$--$20$~MHz. Fitting $\kappa_{\mathrm{bulk}}$ and $f^\ast$ per material ($C$ fixed, $\ell$ derived from $f^\ast$) matches every data set ($\chi^2_\nu<1$); with six points over one decade this is a consistency check of an effective one-length description, not a standalone test of the theory, which rests on the parameter-free calculations of Fig.~\ref{fig:concept} and SM, Figs.~S1--S2. Pure Si and boron-doped Si (Si:B) remain flat ($d_p\gg\ell$; $\ell_{\mathrm{Si}}\lesssim0.6~\mu$m at $1\sigma$), while Ge and the SiGe alloys roll off, yielding $\ell_{\mathrm{Ge}}=0.41\pm0.09$, $\ell_{\mathrm{Si_{0.99}Ge_{0.01}}}=0.34\pm0.07$ and $\ell_{\mathrm{Si_{0.2}Ge_{0.8}}}=0.25\pm0.04~\mu$m ($1\sigma$ statistical; systematic $\pm0.15~\mu$m from the functional form and the treatment of $\kappa_{\mathrm{bulk}}$, SM, Sec.~SVI.A and Table~S1). Within this one-length description the trend across materials is transparent in Eq.~\eqref{eq:Lqq-modal}: mass-disorder scattering~\cite{Koh2007} ($\propto\omega^4$) removes high-frequency modes and \emph{increases} the fraction of $\kappa$ carried by long-MFP modes (from $24\%$ to $47\%$ above $1~\mu$m when $1\%$ Ge is added to Si; SM, Sec.~SVIII), while the lower alloy conductivity shrinks $d_p$ and pushes $k_z=d_p^{-1}$ deeper into the nonlocal regime, so that $\ell$ reflects the MFP window sampled by $d_p(f)$~\cite{Wilson2014,Minnich2015}. Solving the through-plane boundary-value problem directly with the modal kernel (SM, Sec.~SIV.C and Fig.~S2) reproduces $\Lambda_A(f)$ of Ge and Si$_{0.2}$Ge$_{0.8}$ without adjustable parameters under an interface condition adiabatic below 3~THz~\cite{Wilson2013} (Si$_{0.99}$Ge$_{0.01}$ falling between this condition and a fully thermalizing interface); for Si the bulk kernel alone overpredicts the through-plane suppression by an amount set by the fraction of $\kappa$ in long-MFP modes ($19\%$ or $13\%$ decline over the band for $40\%$ or $24\%$ of $\kappa$ above $1~\mu$m, against $7\pm8\%$ measured), the flat $\Lambda_z(f)$ being controlled by the Al/Si interface~\cite{Hua2019} and by the amplitude--phase compensation of the out-of-phase signal~\cite{Koh2014}.

\textit{Role of the interface.---}The same modal picture clarifies why a local boundary conductance is insufficient~\cite{Wilson2014}. Near a metal/sample boundary the reflected and transmitted modes truncate the overlap $\Psi_\mu^{i}(\mathbf{r})\Psi_\nu^{j}(\mathbf{r}')$ within a layer of thickness set by the MFP, leaving residual off-diagonal correlations $R_{\mu\nu}$ ($\mu\neq\nu$) that do not relax to a single local temperature. In the language of Eq.~\eqref{eq:Lqq-modal}, the short- and long-MFP sectors are driven out of mutual equilibrium over a finite length, which is precisely the interfacial nonequilibrium resistance inferred phenomenologically from the frequency dependence of the apparent interface conductance in SiGe alloys~\cite{Wilson2014,Wilson2013}. Projecting Eq.~\eqref{eq:GLE-main} onto a long-MFP sector and a diffusive reservoir recovers the two-channel model of Ref.~\cite{Wilson2013} and maps its nonequilibrium length $d\simeq150$~nm onto the range of the ballistic sector, $\ell_1=d\sqrt{(\Lambda_1+\Lambda_2)/\Lambda_2}\simeq0.28~\mu$m, with $\Lambda_{1,2}$ the conductivities of the two channels (SM, Sec.~SVII). For a thermalizing interface, the boundary-value solution (SM, Sec.~SIV.C) yields the excess resistance $R_{\mathrm{neq}}=\Delta T_{\mathrm{neq}}/q''$, the additional interfacial temperature drop per unit heat flux $q''$ that a local model must assume: $\approx6.5\times10^{-9}$~m$^2$K\,W$^{-1}$ at 1~MHz for Al/Si$_{0.99}$Ge$_{0.01}$ ($3.0$ and $2.5\times10^{-9}$ for Si and Ge). This is comparable to measured interface (Kapitza) resistances ($3$--$7\times10^{-9}$~m$^2$K\,W$^{-1}$ for Al/Si and Al/SiGe~\cite{Minnich2011,Hua2019,Wilson2013}) and decreases with $f$, reproducing the frequency increment, though not the absolute level, of the rise of the apparent interface conductance $G_A(f)$ that accompanies the drop of $\Lambda_A(f)$~\cite{Wilson2013} [SM, Fig.~S2(b)]. The kernel thus places the diffusive reservoir and the nonlocal ballistic channel of the two-fluid description~\cite{Armstrong1981,Maznev2011} on a common, entropy-producing footing, rather than positing them separately.

\textit{Conclusion.---}We have derived a thermodynamically consistent generalization of Fourier's law in which the conductivity is a nonlocal memory operator $\bm{\kappa}_{\mathrm{eff}}(\mathbf{k},\omega)$, whose Onsager kernel decomposes into vibrational normal modes weighted by Bose heat capacities and relaxation functions [Eq.~\eqref{eq:Lqq-modal}]. The kernel is a tensor even in a cubic crystal, and direction-dependent sampling of it yields apparent anisotropy $\Lambda_z\neq\Lambda_r$ in a nominally isotropic crystal; the recognition that TDTR samples $\bm{\kappa}_{\mathrm{eff}}$ at a frequency-set wavevector $k_z\sim d_p^{-1}$ recasts the measured suppression in Ge and SiGe as spatial nonlocality, described by an effective length $\ell\simeq0.25$--$0.4~\mu$m and, without adjustable parameters, by the relaxation-time kernel with the interfacial condition on the long-MFP modes; for the through-plane response of Si the kernel alone overpredicts the suppression and the interface condition dominates; and the truncation of modal overlaps at a boundary provides a microscopic origin for the interfacial nonequilibrium resistance. Together these results place the two-channel ballistic/diffusive picture on a thermodynamic and microscopic footing~\cite{Wilson2014,Armstrong1981,Maznev2011,Wilson2013}, and recover the relaxation-time kernel of the Boltzmann equation as the Markovian closure of the GLE. Relative to BTE-based and hydrodynamic analyses of TDTR~\cite{Koh2014,Hua2019,Ding2014,Xiang2022} the framework adds a second law that holds for any closure consistent with the fluctuation--dissipation theorem, and non-Markovian memory; relative to the Green--Kubo kernel of Ref.~\cite{Zeng2026}, which shares its positivity and its Guyer--Krumhansl limit, it adds the thermodynamic (MNET) route, the fluctuation--dissipation structure of the memory function, and an interface resistance computed from the same kernel, given a modal boundary condition, rather than from a separate interfacial kernel; it does not replace the first-principles input (lifetimes, transmissivities) on which those analyses rely and which it consumes unchanged. Three predictions follow (SM, Sec.~SX): for the same in-plane wavevector the apparent in-plane suppression should exceed the through-plane one by a factor $1.5$--$2$ for $w_0=1$--$10~\mu$m (bounded by 3 at long wavelength) in a material whose through-plane response is not interface-dominated; $\Lambda_r$ measured with different spot sizes and geometries should collapse on the single curve $\kappa_L(k)$ once each geometry is mapped to its effective wavevector by the forward model ($k\simeq2.3/w_0$ at $w_0=1~\mu$m, $3.6/w_0$ at $10~\mu$m); and isotopically enriched $^{28}$Si should show the same $\Lambda_r$ as natural Si at $w_0=1~\mu$m despite an 8\% larger $\kappa_{\mathrm{bulk}}$. Because the kernel is expressed through equilibrium correlation functions and modal heat capacities, the framework connects experimental observables to microscopic parameters without ad hoc extensions, and applies directly to nanowires, superlattices and quantum dots, to the temperature dependence of nondiffusive transport through $c_{\mathrm{ph}}(\omega,T)$, and, by extension to electronic variables, to thermoelectric and active quantum devices~\cite{Gomes-Filho2025}.

\begin{acknowledgments}
The author gratefully acknowledges the financial support by National Institute of Science and Technology in Innovative Research in Health Sciences -- from Nanotechnology to Artificial Intelligence (INCT PICS) sponsored by Brazil’s National Council for Scientific and Technological Development (CNPq), grant no. 408417/2024-2.
\end{acknowledgments}

\bibliography{references_v5}

\end{document}

\title{Supplemental Material for\\``Generalized Fourier's law in mesoscopic systems''}
\author{Luciano C. Lapas\orcidlink{0000-0002-0029-5625}}
\email{luciano.lapas@unila.edu.br}
\affiliation{Latin American Institute of Life and Natural Sciences,
Universidade Federal da Integra\c{c}\~ao Latino-Americana, Av. Tancredo Neves, 6731,
Bloco 6, Espa\c{c}o 3, Sala 5, Foz do Igua\c{c}u, 85867-970, PR, Brazil.}

\maketitle
\beginsupplement

This Supplemental Material provides the derivations summarized in the main text.
Section~\ref{sec:hydro} obtains the nonlocal constitutive relation [Eq.~(2) of the Letter] from entropy production and linear response, and establishes its positive semidefinite (second-law) character.
Section~\ref{sec:meso} traces the kernel to phase-space correlations within mesoscopic nonequilibrium thermodynamics (MNET).
Section~\ref{sec:gle} derives the generalized Langevin equation (GLE) [Eq.~(4)] by Mori--Kubo--Zwanzig projection.
Section~\ref{sec:modal} performs the modal decomposition leading to the closed-form, tensorial Onsager kernel [Eq.~(6)] and its quantum (Bose) weighting [Eq.~(7)].
Section~\ref{sec:spectral} classifies the transport regimes from the low-frequency memory spectrum.
Section~\ref{sec:aniso} collects technical notes on anisotropy, spatial-wavevector nonlocality, and the interface connection.
Section~\ref{sec:rta} derives the relaxation-time closure of the kernel and its longitudinal/transverse structure; Sec.~\ref{sec:bvp} solves the through-plane boundary-value problem that yields the frequency dependence of the apparent conductivity and interface conductance; Sec.~\ref{sec:twosector} shows that the two-channel model is a two-sector projection of the GLE; Sec.~\ref{sec:spectra} describes the mean-free-path spectra and the TDTR forward model used in Figs.~S1--S2; Sec.~\ref{sec:relation} compares the framework item by item with previous nonlocal treatments; Sec.~\ref{sec:pred} collects testable predictions.

\emph{Reader's guide and consistency checks.}---The chain of assumptions is: (1) Gibbs equation extended to phase-space variables and linear response consistent with $\sigma\ge0$ [Secs.~\ref{sec:hydro}--\ref{sec:meso}; nothing microscopic is assumed]; (2) the Mori--Kubo--Zwanzig identity, an exact operator identity for the Liouville dynamics [Sec.~\ref{sec:gle}; standard, see Zwanzig~\cite{Zwanzig2001}, Ch.~8]; (3) the choice of slow variables as modal energies, which gives Eq.~(6) of the Letter without further approximation but leaves $R_{\mu\nu}$ unspecified [Sec.~\ref{sec:modal}]; (4) a closure of the memory kernel, Markovian and diagonal (relaxation-time approximation) for all numerical results [Sec.~\ref{sec:rta}]; (5) the boundary condition at the interface, in two limiting forms [Sec.~\ref{sec:bvp}]. Only steps (4)--(5) are approximations in the physical sense; what they resolve, replace or omit is listed in Sec.~\ref{sec:scope}. The following checks are made explicitly: dimensional consistency of Eq.~(6) [after Eq.~\eqref{eq:Lqq-quantum}]; the $\mathbf{k}\to0$ limit of the closed kernel reproduces the Green--Kubo bulk conductivity $\sum_\mu C_\mu v_\mu^2\tau_\mu/3V$ [Eq.~\eqref{eq:rta-k}]; the Markovian closure gives Fourier's law for $\omega\tau\ll1$ and, when its single inertial pole is retained, the Cattaneo--Vernotte and Guyer--Krumhansl forms [Sec.~\ref{sec:spectral}]; the relaxation-time kernel coincides with the published Boltzmann result [Eq.~\eqref{eq:rta-k} vs.\ Eq.~(22) of Ref.~\cite{Hua2019}]; the angular averages $F_L$, $F_T$ have the expansions and tails stated in Eqs.~\eqref{eq:FL}--\eqref{eq:FT} (verified by quadrature); the numerical boundary-value solver recovers the Fourier limit to $10^{-4}$ and the analytic injection resistance of a single ballistic bin [Sec.~\ref{sec:bvp}]. Throughout, $\mu=(\mathbf{q},s)$ labels a phonon mode of wavevector $\mathbf{q}$ and branch $s$, with $\omega_\mu=\omega_s(\mathbf{q})$, $\mathbf{v}_\mu=\nabla_{\mathbf{q}}\omega_s(\mathbf{q})$ and $\tau_\mu=\tau_s(\mathbf{q})$; $\mathbf{k}$ is the wavevector of the temperature field and is distinct from $\mathbf{q}$.

% ======================================================================
\suppsection{Hydrodynamic derivation and second law}{sec:hydro}

At the hydrodynamic level the total entropy is the volume integral of the local entropy density $s(\mathbf{r},t)$~\cite{deGroot1984},
\begin{equation}
S(t)=\int d\mathbf{r}\,s(\mathbf{r},t),\qquad
\frac{dS}{dt}=\int d\mathbf{r}\,\sigma(\mathbf{r},t)\ge 0,
\label{eq:total-entropy}
\end{equation}
where $\sigma$ is the local production rate. For heat transport in a single-component rigid medium (no mass flow), the entropy balance gives the bilinear form $\sigma(\mathbf{r},t)=\mathbf{J}_q(\mathbf{r},t)\cdot\mathbf{X}_q(\mathbf{r},t)\ge0$, identifying the conjugate force
\begin{equation}
\mathbf{X}_q(\mathbf{r},t)\equiv\nabla\!\left[\frac{1}{T(\mathbf{r},t)}\right].
\label{eq:force}
\end{equation}
For small deviations from equilibrium the flux is a linear functional of the force history~\cite{Onsager1931a,Onsager1931b}. The most general causal, time-translation-invariant linear law is
\begin{equation}
\mathbf{J}_q(\mathbf{r},t)=\int_{-\infty}^{t}\!dt'\!\int d\mathbf{r}'\,
\mathbf{L}_{qq}(\mathbf{r},\mathbf{r}',t-t')\,\mathbf{X}_q(\mathbf{r}',t'),
\label{eq:nonlocal}
\end{equation}
with $\mathbf{L}_{qq}=\bm{\Lambda}_{qq}\,\Theta(t-t')$ and $\Theta$ the Heaviside function. Equation~\eqref{eq:nonlocal} is Eq.~(2) of the Letter. Inserting it into Eq.~\eqref{eq:total-entropy} yields the quadratic functional
\begin{equation}
\frac{dS}{dt}=\int_{-\infty}^{t}\!dt'\!\iint d\mathbf{r}\,d\mathbf{r}'\,
\mathbf{X}_q(\mathbf{r},t)\cdot\mathbf{L}_{qq}(\mathbf{r},\mathbf{r}',t-t')\cdot\mathbf{X}_q(\mathbf{r}',t'),
\label{eq:Sdot}
\end{equation}
which must be nonnegative for arbitrary small histories of $\mathbf{X}_q$. This implies that the symmetric part of $\mathbf{L}_{qq}$ defines a positive semidefinite operator (generalized Onsager positivity condition for memory kernels). For stationary kernels it is convenient to state this condition in the Fourier/Laplace domain: the dissipative (Hermitian) part of $\widetilde{\mathbf{L}}_{qq}(\mathbf{k},\omega)$ must be positive semidefinite, i.e.\ $\mathrm{Re}\,\widetilde{\mathbf{L}}_{qq}(\mathbf{k},\omega)\succeq 0$ for all $\mathbf{k}$ and real $\omega$.

% ======================================================================
\suppsection{Mesoscopic origin of the kernel}{sec:meso}

Within MNET~\cite{Vilar2001,Reguera2005}, thermal transport is the stochastic evolution of carriers (phonons, photons, quasiparticles) over a phase-space landscape $\bm{\Gamma}=(\mathbf{p},\mathbf{r})$ with density $\rho(\bm{\Gamma},t)$. The Gibbs entropy postulate $S=-k_B\!\int\rho\ln(\rho/\rho_{\mathrm{leq}})\,d\bm{\Gamma}+S_{\mathrm{leq}}$, together with the phase-space continuity equation $\partial_t\rho=-\nabla_{\bm{\Gamma}}\cdot\mathbf{J}$, gives the production rate
\begin{equation}
\sigma=-\int \mathbf{J}(\bm{\Gamma},t)\cdot\nabla_{\bm{\Gamma}}\!\left[\frac{\mu(\bm{\Gamma},t)}{T}\right]d\bm{\Gamma}\ge0,
\quad
\mu=k_BT\ln\frac{\rho}{\rho_{\mathrm{leq}}}+\mu_{\mathrm{leq}}.
\label{eq:mu}
\end{equation}
A causal linear law with a nonlocal phase-space kernel $\mathbf{L}(\bm{\Gamma},\bm{\Gamma}',t-t')$,
\begin{equation}
\mathbf{J}(\bm{\Gamma},t)=\int_{-\infty}^{t}\!dt'\!\int d\bm{\Gamma}'\,
\mathbf{L}(\bm{\Gamma},\bm{\Gamma}',t-t')\,\nabla_{\bm{\Gamma}'}\!\left[\frac{\mu(\bm{\Gamma}',t')}{T}\right],
\label{eq:J-L-mu}
\end{equation}
recovers the standard Fokker--Planck equation in the Markovian limit $\mathbf{L}\propto\delta(t)$. Under local quasi-equilibrium the chemical-potential gradient is dominated by the thermal gradient, $\nabla_{\bm{\Gamma}'}(\mu/T)\simeq\mathcal{A}(\bm{\Gamma}';T)\,\nabla_{\mathbf{r}'}(1/T)$. Coarse-graining the spatial current $\mathbf{J}_q(\mathbf{r},t)=\int d\bm{\Gamma}\,\varepsilon(\bm{\Gamma})\mathbf{J}_{\mathbf{r}}(\bm{\Gamma},t)\,\delta(\mathbf{r}-\mathbf{r}(\bm{\Gamma}))$ then reproduces Eq.~\eqref{eq:nonlocal} with
\begin{equation}
\mathbf{L}_{qq}(\mathbf{r},\mathbf{r}',\tau)=\iint d\bm{\Gamma}\,d\bm{\Gamma}'\,
\varepsilon(\bm{\Gamma})\,\mathbf{L}_{\mathbf{rr}}(\bm{\Gamma},\bm{\Gamma}',\tau)\,\mathcal{A}(\bm{\Gamma}')\,
\delta(\mathbf{r}-\mathbf{r}(\bm{\Gamma}))\,\delta(\mathbf{r}'-\mathbf{r}(\bm{\Gamma}')).
\label{eq:Lqq-coarse}
\end{equation}
Spatial nonlocality ($\mathbf{r}\neq\mathbf{r}'$) thus originates in phase-space correlations that couple distinct positions through carrier streaming and scattering.

% ======================================================================
\suppsection{Generalized Langevin equation}{sec:gle}

Let $\mathcal{L}$ be the Liouville operator, $\dot{X}=i\mathcal{L}X$, and let $\mathbf{A}(\bm{\Gamma})$ be the column vector of relevant slow variables (energy density and heat-flux modes). The Mori--Zwanzig projector and its complement are
\begin{equation}
\mathcal{P}Y=\langle Y\mathbf{A}^{\dagger}\rangle\,\mathbf{C}^{-1}\,\mathbf{A},
\qquad \mathcal{Q}=1-\mathcal{P},
\qquad \mathbf{C}=\langle\mathbf{A}\mathbf{A}^{\dagger}\rangle,
\label{eq:proj}
\end{equation}
with the inner product $(X,Y)=\langle XY^{\dagger}\rangle_{\rho_{\mathrm{eq}}}$. The operator identity
$e^{it\mathcal{L}}=e^{it\mathcal{Q}\mathcal{L}}+\int_0^t d\tau\,e^{i(t-\tau)\mathcal{L}}\,i\mathcal{P}\mathcal{L}\,e^{i\tau\mathcal{Q}\mathcal{L}}$
applied to $\dot{\mathbf{A}}=e^{it\mathcal{L}}i\mathcal{L}\mathbf{A}$ gives the GLE
\begin{equation}
\dot{\mathbf{A}}(t)=i\bm{\Omega}\,\mathbf{A}(t)-\int_0^t\!d\tau\,\bm{\Phi}(\tau)\,\mathbf{A}(t-\tau)+\mathbf{F}(t),
\label{eq:GLE}
\end{equation}
which is Eq.~(4) of the Letter, with
\begin{equation}
\bm{\Omega}=\langle(i\mathcal{L}\mathbf{A})\mathbf{A}^{\dagger}\rangle\,\mathbf{C}^{-1},
\quad
\mathbf{F}(t)=e^{it\mathcal{Q}\mathcal{L}}\,\mathcal{Q}\,(i\mathcal{L}\mathbf{A}),
\quad
\bm{\Phi}(\tau)=\langle\mathbf{F}(\tau)\mathbf{F}^{\dagger}(0)\rangle\,\mathbf{C}^{-1}.
\label{eq:GLE-defs}
\end{equation}
The last relation is the second fluctuation--dissipation theorem; the projected propagator $e^{it\mathcal{Q}\mathcal{L}}$ keeps $\mathbf{F}(t)$ orthogonal to $\mathbf{A}(0)$. For real variables even under time reversal, $\bm{\Omega}=\mathbf{0}$. This is not the case for the complex Fourier components of the modal energy densities used in Sec.~\ref{sec:rta}: time reversal maps $\mu=(\mathbf{q},s)$ onto $\bar\mu=(-\mathbf{q},s)$ with $\mathbf{v}_{\bar\mu}=-\mathbf{v}_\mu$, and $\langle(i\mathcal{L}\,\delta E_\mu(\mathbf{k}))\,\delta E_\mu^{\dagger}(\mathbf{k})\rangle=-i\,\mathbf{k}\cdot\mathbf{v}_\mu\langle|\delta E_\mu(\mathbf{k})|^2\rangle\neq0$; the frequency matrix $i\Omega_{\mu\nu}(\mathbf{k})=-i\,\mathbf{k}\cdot\mathbf{v}_\mu\delta_{\mu\nu}$ is the free-streaming term and carries the spatial nonlocality of a bulk crystal. Laplace transforming ($\widetilde{f}(z)=\int_0^\infty e^{-zt}f\,dt$),
\begin{equation}
\widetilde{\mathbf{A}}(z)=\widetilde{\mathbf{R}}(z)\,\mathbf{A}(0)+\widetilde{\mathbf{R}}(z)\,\widetilde{\mathbf{F}}(z),
\qquad
\widetilde{\mathbf{R}}(z)=[z\mathbf{1}-i\bm{\Omega}+\widetilde{\bm{\Phi}}(z)]^{-1}.
\label{eq:R-laplace}
\end{equation}
Right-multiplying by $\mathbf{A}^{\dagger}(0)$ and averaging removes the noise term (orthogonality), giving $\mathbf{C}(t)=\mathbf{R}(t)\mathbf{C}_0$ with $\mathbf{R}(0)=\mathbf{1}$, so that $\mathbf{R}(t)=\mathbf{C}(t)\mathbf{C}_0^{-1}$ governs the decay of fluctuations. The decay rates are the poles of $\widetilde{\mathbf{R}}(z)$, i.e.\ the roots of $\det[z\mathbf{1}-i\bm{\Omega}+\widetilde{\bm{\Phi}}(z)]=0$; for the relaxation-time closure of Sec.~\ref{sec:rta}, $-i\bm{\Omega}$ supplies the streaming term $i\mathbf{k}\cdot\mathbf{v}_\mu$ in Eq.~\eqref{eq:rta-kw}.

% ======================================================================
\suppsection{Modal decomposition and quantum weighting}{sec:modal}

In a structure with dimensions comparable to the phonon MFP, the continuum spectrum is replaced by discrete eigenmodes $\phi_\mu(\mathbf{r})$ satisfying $\hat{\mathcal{L}}_{\mathrm{wave}}\phi_\mu=-\omega_\mu^2\phi_\mu$ in $V$, orthonormalized as $\int_V\phi_\mu^*\phi_\nu\,d\mathbf{r}=\delta_{\mu\nu}$. The harmonic energy density is $u(\mathbf{r},t)=\sum_\mu\varepsilon_\mu n_\mu(t)|\phi_\mu(\mathbf{r})|^2$, and the kinetic (diagonal) heat flux is
\begin{equation}
\mathbf{J}_q(\mathbf{r},t)=\sum_\mu\bm{\Psi}_\mu(\mathbf{r})\,A_\mu(t),
\qquad
\bm{\Psi}_\mu(\mathbf{r})=\mathbf{v}_\mu(\mathbf{r})\,|\phi_\mu(\mathbf{r})|^2,
\label{eq:modal-flux}
\end{equation}
with $A_\mu(t)=\delta E_\mu(t)=\varepsilon_\mu[n_\mu(t)-n_\mu^{\mathrm{eq}}]$ the modal energy fluctuation, $\varepsilon_\mu=\hbar\omega_\mu$, and $\mathbf{v}_\mu$ the local group velocity. Taking the slow variable to be the modal energy rather than the bare population keeps the heat-capacity weighting consistent below. Inserting Eq.~\eqref{eq:modal-flux} into the Green--Kubo definition of the Onsager coefficient conjugate to $\nabla(1/T)$, $L_{qq}^{ij}(\mathbf{r},\mathbf{r}',t)=k_B^{-1}\langle J_q^{i}(\mathbf{r},t)J_q^{j}(\mathbf{r}',0)\rangle$, so that $\kappa^{ij}=L_{qq}^{ij}/T^2$ is the spatiotemporal Green--Kubo kernel of Ref.~\cite{Zeng2026} and using $\mathbf{C}(t)=\mathbf{R}(t)\mathbf{C}_0$ with diagonal static fluctuations $\langle A_\mu^2\rangle_{\mathrm{eq}}=\langle\delta E_\mu^2\rangle_{\mathrm{eq}}$ gives
\begin{equation}
L_{qq}^{ij}(\mathbf{r},\mathbf{r}',t)=\frac{1}{k_B}\sum_{\mu,\nu}\Psi_\mu^{i}(\mathbf{r})\,\Psi_\nu^{j}(\mathbf{r}')\,R_{\mu\nu}(t)\,\langle A_\nu^2\rangle_{\mathrm{eq}}.
\label{eq:Lqq-classical}
\end{equation}
For bosonic carriers the population variance is $\langle(\delta n_\nu)^2\rangle_{\mathrm{eq}}=n_\nu^{\mathrm{eq}}(n_\nu^{\mathrm{eq}}+1)$, so the energy variance equals $\langle\delta E_\nu^2\rangle=(\hbar\omega_\nu)^2\langle(\delta n_\nu)^2\rangle_{\mathrm{eq}}=k_BT^2\,c_{\mathrm{ph}}(\omega_\nu,T)$, with
\begin{equation}
c_{\mathrm{ph}}(\omega,T)=k_B\!\left(\frac{\hbar\omega}{2k_BT}\right)^2\!\mathrm{csch}^2\!\left(\frac{\hbar\omega}{2k_BT}\right).
\label{eq:cph-SM}
\end{equation}
The prefactor $k_B$ cancels, yielding the quantum-corrected tensorial Onsager kernel [Eq.~(6) of the Letter],
\begin{equation}
L_{qq}^{ij}(\mathbf{r},\mathbf{r}',t)=T^2\sum_{\mu,\nu}\Psi_\mu^{i}(\mathbf{r})\,\Psi_\nu^{j}(\mathbf{r}')\,R_{\mu\nu}(t)\,c_{\mathrm{ph}}(\omega_\nu,T).
\label{eq:Lqq-quantum}
\end{equation}
so that $\kappa^{ij}(\mathbf{r},\mathbf{r}',t)=L_{qq}^{ij}/T^2=\sum_{\mu\nu}\Psi^i_\mu\Psi^j_\nu R_{\mu\nu}c_{\mathrm{ph}}(\omega_\nu,T)$; dimensionally, $[\Psi]=\mathrm{m^{-2}s^{-1}}$ and $[c_{\mathrm{ph}}]=\mathrm{J\,K^{-1}}$ give $[\kappa(\mathbf{r},\mathbf{r}',t)]=\mathrm{W\,K^{-1}m^{-4}s^{-1}}$, as required by Eq.~\eqref{eq:nonlocal}. For extended modes the mode sum is per unit volume, $V^{-1}\sum_\mu$.

\emph{Second law.}---Defining projected forces $X_\mu(t)=-T_R^{-1}\!\int_V d\mathbf{r}\,\bm{\Psi}_\mu(\mathbf{r})\cdot\nabla T(\mathbf{r},t)$, the entropy production [Eq.~\eqref{eq:Sdot}] becomes the quadratic form $\sigma[T]=(k_BT_R^2)^{-1}\sum_{\mu,\nu}\int_{-\infty}^t dt'\,X_\mu(t)\,C_{\mu\nu}(t-t')\,X_\nu(t')$. Since $C_{\mu\nu}(t)$ is a stationary correlation matrix, the Wiener--Khinchin and Bochner theorems guarantee a non-negative-definite power spectrum, hence $\sigma[T]\ge0$. The argument holds for any closure of $\bm{\Phi}$ that keeps $C_{\mu\nu}(t)$ the correlation matrix of a stationary process, i.e.\ any closure consistent with the fluctuation--dissipation theorem of Eq.~\eqref{eq:GLE-defs} (non-negative memory spectrum, as for the relaxation-time closure of Sec.~\ref{sec:rta}); an arbitrary truncation of $\widetilde{\bm{\Phi}}(z)$ need not satisfy it.

\suppsubsection{Long-wavelength expansion and a minimal one-length $k$-space form}{sec:lw-expansion}

This subsection provides a short derivation of the minimal Lorentzian wavevector dependence used in the Letter,
$\kappa_{\mathrm{eff}}(k)=\kappa_{\mathrm{bulk}}/[1+(k\ell)^2]$, as a controlled long-wavelength approximation.

\emph{Quasistatic reduction to a spatial kernel.}---In the quasistatic regime relevant to the MHz TDTR fits in the Letter,
the dominant effect is spatial nonlocality and one may consider the time-integrated (zero-frequency) kernel
\begin{equation}
\overline{L}_{qq}^{ij}(\mathbf{r},\mathbf{r}') \equiv \int_{0}^{\infty} dt\, L_{qq}^{ij}(\mathbf{r},\mathbf{r}',t),
\label{eq:Lbar_def}
\end{equation}
and the corresponding spatial conductivity kernel
$\overline{\kappa}^{ij}(\mathbf{r},\mathbf{r}')=\overline{L}_{qq}^{ij}(\mathbf{r},\mathbf{r}')/T_0^2$.

\emph{Isotropic reduction and analyticity at $k=0$.}---Assume an isotropic medium and focus on a diagonal component. Writing the spatially homogeneous kernel as an even function of the separation,
\begin{equation}
\overline{\kappa}(\mathbf{r},\mathbf{r}') \equiv \overline{\kappa}(\boldsymbol{\xi}),\qquad \boldsymbol{\xi}\equiv \mathbf{r}-\mathbf{r}',\qquad
\overline{\kappa}(\boldsymbol{\xi})=\overline{\kappa}(-\boldsymbol{\xi}),
\end{equation}
its spatial Fourier transform is real and analytic near $k=0$:
\begin{equation}
\kappa(k)\equiv \int d\boldsymbol{\xi}\,e^{-i\mathbf{k}\cdot\boldsymbol{\xi}}\overline{\kappa}(\boldsymbol{\xi})
= \kappa(0)\left[1-a k^2 + O(k^4)\right].
\label{eq:kappa-exp}
\end{equation}
The coefficient $a$ is fixed by the second spatial moment of the real-space kernel. Expanding the exponential and using isotropy,
\begin{align}
\kappa(k)
&= \int d\boldsymbol{\xi}\,\overline{\kappa}(\boldsymbol{\xi})
-\frac{1}{2}\sum_{m,n}k_m k_n \int d\boldsymbol{\xi}\,\xi_m\xi_n\,\overline{\kappa}(\boldsymbol{\xi})
+O(k^4)\nonumber\\
&= \kappa(0)-\frac{k^2}{2d}\int d\boldsymbol{\xi}\,\xi^2\,\overline{\kappa}(\boldsymbol{\xi})+O(k^4),
\end{align}
where $d$ is the spatial dimension. Thus,
\begin{equation}
a=\frac{1}{2d}\,\frac{\int d\boldsymbol{\xi}\,\xi^2\,\overline{\kappa}(\boldsymbol{\xi})}{\int d\boldsymbol{\xi}\,\overline{\kappa}(\boldsymbol{\xi})}.
\label{eq:alpha-moment}
\end{equation}
This identifies a natural nonlocal length scale,
\begin{equation}
\ell^2 \equiv a,
\label{eq:ell-def}
\end{equation}
i.e.\ $\ell$ is the rms range of the conductivity kernel (up to the geometric factor $2d$).

\emph{Minimal Pad\'e resummation.}---A convenient closed form that (i) matches Eq.~\eqref{eq:kappa-exp} to order $k^2$, (ii) is positive and monotonically decreasing with $k^2$, and (iii) corresponds to a finite-range real-space kernel, is the $[0/1]$ Pad\'e approximant
\begin{equation}
\kappa(k)\simeq \frac{\kappa(0)}{1+\ell^2 k^2}.
\label{eq:pade}
\end{equation}

\begin{figure}[t]
\centering
\includegraphics[width=\linewidth]{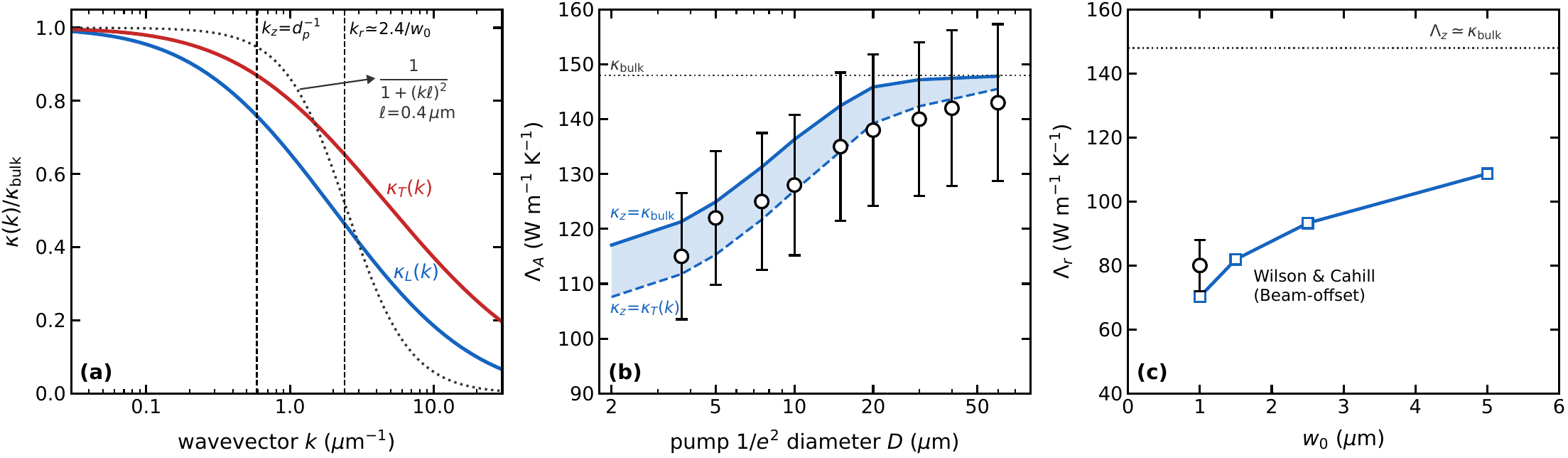}
\caption{(a) Longitudinal and transverse blocks of the relaxation-time kernel of Si at 300~K (also Fig.~1 of the Letter), Eqs.~\eqref{eq:FL}--\eqref{eq:FT}, from the MFP spectrum of Sec.~\ref{sec:spectra} (anchored to published first-principles and MFP-spectroscopy values); the one-length Lorentzian with $\ell=0.4~\mu$m (dotted) does not represent the kernel in the TDTR window. Dashed lines: through-plane ($k_z=d_p^{-1}$ at 9.8~MHz) and in-plane ($k_r\simeq2.4/w_0$, $w_0=1~\mu$m) wavevectors. (b) Apparent conductivity from concentric TDTR versus pump $1/e^2$ diameter obtained by inserting the kernel in the thermal model of the experiment and fitting the synthetic signal with the local model (solid: $\kappa_z=\kappa_{\mathrm{bulk}}$; dashed: $\kappa_z=\kappa_T(k)$), compared with Ding \emph{et al.}~\cite{Ding2014} (symbols, 10\% error bars). (c) In-plane conductivity from the beam-offset width versus spot size (squares) and the measurement of Wilson and Cahill~\cite{Wilson2014}. No parameter is adjusted to the data.}
\label{fig:S1}
\end{figure}

\suppsubsection{Relaxation-time closure and tensor structure}{sec:rta}
For a bulk crystal the modes are Bloch waves with $|\phi_\mu|^2=1/V$, so $\widetilde{\bm{\Psi}}_\mu(\mathbf{k})=\mathbf{v}_\mu\delta_{\mathbf{k},0}$ and the overlap in Eq.~\eqref{eq:Lqq-quantum} carries no $\mathbf{k}$ dependence; nonlocality resides in the propagator. Taking as slow variables the local modal energy densities $A_\mu(\mathbf{k},t)$, the frequency matrix of Eq.~\eqref{eq:GLE} is $i\Omega_{\mu\nu}=-i\mathbf{k}\cdot\mathbf{v}_\mu\delta_{\mu\nu}$ (Sec.~\ref{sec:gle}). The simplest closure of the dissipative part is Markovian and diagonal, $\Phi_{\mu\nu}(t)=2\tau_\mu^{-1}\delta_{\mu\nu}\delta(t)$ (relaxation-time approximation, RTA), giving
\begin{equation}
R_{\mu\nu}(\mathbf{k},t)=\delta_{\mu\nu}\exp\!\big[-t/\tau_\mu-i\,\mathbf{k}\cdot\mathbf{v}_\mu t\big].
\end{equation}
With $J^i_q(\mathbf{k},t)=\sum_\mu v^i_\mu A_\mu(\mathbf{k},t)$ and $\langle|A_\mu|^2\rangle=k_BT^2c_{\mathrm{ph}}(\omega_\mu,T)$, the spatiotemporal Green--Kubo relation yields
\begin{equation}
\kappa^{ij}(\mathbf{k},\omega)=\frac{1}{k_BT^2V}\int_0^\infty\!dt\,e^{-i\omega t}\langle J^i_q(\mathbf{k},t)J^j_q(-\mathbf{k},0)\rangle
=\frac1V\sum_\mu\frac{c_{\mathrm{ph}}(\omega_\mu,T)\,v^i_\mu v^j_\mu\tau_\mu}{1+i\omega\tau_\mu+i\,\mathbf{k}\cdot\bm{\Lambda}_\mu},\qquad \bm{\Lambda}_\mu=\mathbf{v}_\mu\tau_\mu,
\label{eq:rta-kw}
\end{equation}
and, combining the pair $(\mu,\bar\mu)$ at $\omega=0$,
\begin{equation}
\kappa^{ij}(\mathbf{k})=\frac1V\sum_\mu\frac{C_\mu\,v^i_\mu v^j_\mu\tau_\mu}{1+(\mathbf{k}\cdot\bm{\Lambda}_\mu)^2},\qquad
C_\mu\equiv c_{\mathrm{ph}}(\omega_\mu,T)=\hbar\omega_\mu\,\frac{\partial n_\mu^{\mathrm{eq}}}{\partial T},
\label{eq:rta-k}
\end{equation}
the nonlocal conductivity of the Peierls--Boltzmann equation in the RTA~\cite{Mahan1988,Koh2014,Allen2018,Hua2019}; the Bose weight of Eq.~\eqref{eq:cph-SM} is the modal heat capacity of kinetic theory. The GLE thus contains the kinetic kernel as its Markovian closure, while admitting non-Markovian and intermode memory (Sec.~\ref{sec:spectral}) and guaranteeing $\sigma\ge0$ for every closure consistent with the fluctuation--dissipation theorem. For an isotropic distribution of $\hat{\mathbf{v}}_\mu$ at fixed $\Lambda$, write $\mathbf{k}\cdot\bm{\Lambda}=k\Lambda\mu$, $\mu=\cos\theta$. The static kernel is then a tensor even in a cubic crystal, $\kappa^{ij}(\mathbf{k})=\kappa_L(k)\hat k_i\hat k_j+\kappa_T(k)(\delta_{ij}-\hat k_i\hat k_j)$, with $\kappa_{L,T}(k)=\sum_\mu\kappa_\mu F_{L,T}(k\Lambda_\mu)$, $\kappa_\mu=C_\mu v_\mu^2\tau_\mu/3V$, and
\begin{align}
F_L(x)&=\int_0^1\!\frac{3\mu^2\,d\mu}{1+x^2\mu^2}=\frac{3(x-\arctan x)}{x^3}=1-\frac35x^2+\frac37x^4-\dots\;\xrightarrow{\;x\gg1\;}\;\frac{3}{x^2},
\label{eq:FL}\\
F_T(x)&=\int_0^1\!\frac{\tfrac32(1-\mu^2)\,d\mu}{1+x^2\mu^2}=\frac32\!\left[\frac{(1+x^2)\arctan x}{x^3}-\frac1{x^2}\right]=1-\frac15x^2+\frac{3}{35}x^4-\dots\;\xrightarrow{\;x\gg1\;}\;\frac{3\pi}{4x}.
\label{eq:FT}
\end{align}
A gradient along the probed wavevector is suppressed three times more strongly than a gradient transverse to it, and the transverse block decays more slowly. Matching the $O(k^2)$ terms to the Pad\'e form of Sec.~\ref{sec:lw-expansion} identifies
\begin{equation}
\ell_L^2=\tfrac35\langle\Lambda^2\rangle_\kappa,\qquad\ell_T^2=\tfrac15\langle\Lambda^2\rangle_\kappa,\qquad
\langle\Lambda^2\rangle_\kappa=\frac{\sum_\mu\kappa_\mu\Lambda_\mu^2}{\sum_\mu\kappa_\mu}=\int\Lambda^2\,d\alpha(\Lambda),
\label{eq:ellLT}
\end{equation}
with $\alpha(\Lambda)$ the conductivity accumulation function~\cite{Minnich2011}. The nonlocality length of the one-length form is therefore the $\kappa$-weighted r.m.s.\ mean free path, dominated by the longest MFPs; for Si at 300~K the spectrum of Sec.~\ref{sec:spectra} gives $\ell_L\approx5~\mu$m and the Pad\'e form is useful only for $k\lesssim0.2~\mu$m$^{-1}$, whereas the full sums decay almost logarithmically in $k$ across the TDTR window [Fig.~\ref{fig:S1}(a), reproduced as Fig.~1 of the Letter]. A Lorentzian fitted to the full $\kappa_L(k)$ within $k=0.3$--$1.2~\mu$m$^{-1}$ returns $\ell_{\mathrm{eff}}\approx0.6~\mu$m and $\kappa(0)\approx0.8\,\kappa_{\mathrm{bulk}}$, which is the status of the parameters fitted in Fig.~2 of the Letter. Equation~\eqref{eq:rta-k} coincides with the transient-grating conductivity of Ref.~\cite{Hua2019} [their Eq.~(22)] and with the suppression-function formulation of MFP spectroscopy~\cite{Minnich2011,Ding2014}, $\Lambda_A=\int S(\Lambda/L)\,d\alpha(\Lambda)$ with $S=F_L$ for a sinusoidal in-plane gradient.

\begin{figure}[t]
\centering
\includegraphics[width=0.95\linewidth]{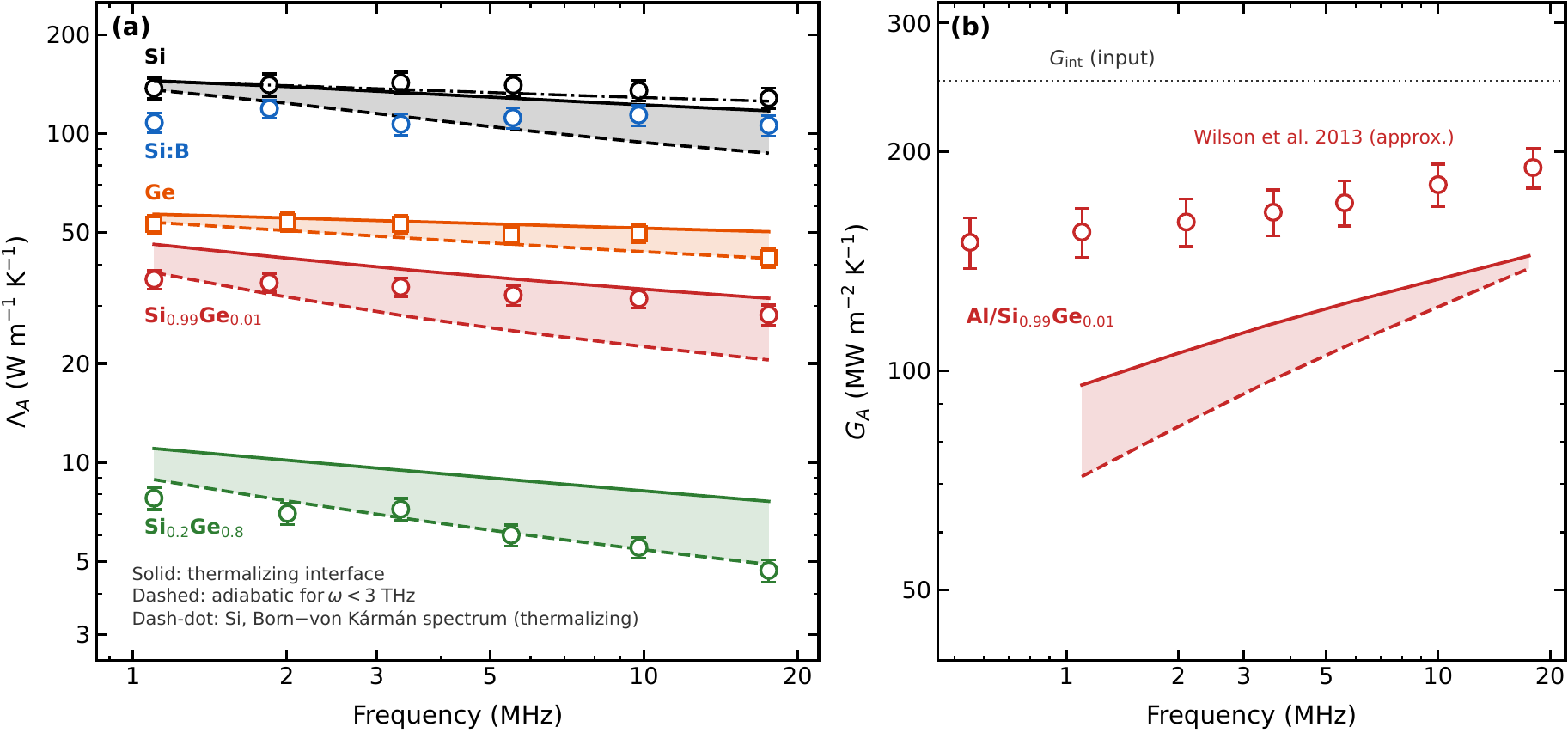}
\caption{(a) Apparent thermal conductivity versus modulation frequency from the one-dimensional modal-kernel calculation of Sec.~\ref{sec:bvp} inserted in the TDTR thermal model, for a thermalizing interface (solid) and for modes below 3~THz adiabatic (dashed; for Si, modes with $\Lambda>0.7~\mu$m); dash-dotted: Si with the Born--von K\'arm\'an spectrum of Sec.~\ref{sec:spectra} and a thermalizing interface; against the data of Wilson and Cahill~\cite{Wilson2014} (symbols, $w_0=10~\mu$m; no model is shown for Si:B). No parameter is adjusted to the data. (b) Apparent interface conductance of Al/Si$_{0.99}$Ge$_{0.01}$ from the same calculation with $G_{\mathrm{int}}=250$~MW\,m$^{-2}$K$^{-1}$ (dotted) and the measurement of Wilson \emph{et al.}~\cite{Wilson2013} (symbols, read from their Fig.~3).}
\label{fig:S2}
\end{figure}

\suppsubsection{Through-plane response: boundary-value problem}{sec:bvp}
For the through-plane direction of a semi-infinite sample the kernel alone does not fix the response: each modal stream requires a boundary condition at $z=0$. Writing $g^\pm_\mu(z)$ for the forward/backward deviational energy densities of mode $\mu$ (direction cosine $\mu_z>0$), the RTA streams under periodic heating $e^{i\omega t}$ are [Eqs.~(6)--(13) of Ref.~\cite{Hua2019}]
\begin{equation}
g^+_\mu(z)=P_\mu e^{-z/\lambda_\mu}+\frac{C_\mu}{\lambda_\mu\alpha_\mu}\!\int_0^z\!e^{-(z-z')/\lambda_\mu}T(z')\,dz',\qquad
g^-_\mu(z)=\frac{C_\mu}{\lambda_\mu\alpha_\mu}\!\int_z^\infty\!e^{-(z'-z)/\lambda_\mu}T(z')\,dz',
\label{eq:streams}
\end{equation}
with $\alpha_\mu=1+i\omega\tau_\mu$ and $\lambda_\mu=\Lambda_\mu\mu_z/\alpha_\mu$; the local temperature is fixed by energy conservation, $\sum_\mu\tau_\mu^{-1}[g^+_\mu+g^-_\mu-2C_\mu T]=0$, and the surface flux is $q_0=\sum_\mu v_\mu\mu_z[g^+_\mu(0)-g^-_\mu(0)]$. The constant $P_\mu$ encodes the interface and is the quantity that distinguishes materials with the same bulk kernel: for a \emph{thermalizing} interface $P_\mu=C_\mu T_w$, with $T_w$ the interface temperature (every mode is emitted at $T_w$, the $R=0$ limit of Ref.~\cite{Wilson2013}); for an \emph{adiabatic} mode $P_\mu=g^-_\mu(0)$ (specular reflection, no net flux, the $R=1$ limit). Following Ref.~\cite{Wilson2013}, the adiabatic condition is applied to modes below 3~THz in the second case (for Si, whose spectrum is resolved in MFP only, to the bins with $\Lambda>0.7~\mu$m). Modes with $\Lambda_\mu<30$~nm are treated as a Fourier reservoir together with the optical heat capacity. Equations~\eqref{eq:streams} are evaluated after two integrations by parts, $g^+_\mu+g^-_\mu=(C_\mu/\alpha_\mu)[2T+\lambda_\mu\!\int e^{-|z-z'|/\lambda_\mu}T''(z')dz'+\text{boundary terms}]$, which makes the local limit exact on a nonuniform grid (260 points from 0.3~nm to $14d_p$, 26 MFP bins, 8-point Gauss--Legendre quadrature in $\mu_z$); the Fourier limit is recovered to $10^{-4}$ and the ballistic injection resistance of a single long-MFP bin, $4/(C_bv_b)$, is reproduced analytically. The surface impedance $Z(\omega)=T_w/q_0$ defines a complex through-plane conductivity $\kappa_z(\omega)=1/(i\omega CZ^2)$, tabulated at 22 frequencies between 0.3~MHz and 20~GHz and inserted as the substrate conductivity in the multilayer Hankel-space model of Sec.~\ref{sec:spectra}. The synthetic in-phase/out-of-phase ratio over 0.1--4~ns is fitted with the local model for the pair $(\Lambda_A,G_A)$, exactly as in Refs.~\cite{Wilson2014,Wilson2013}. Figure~\ref{fig:S2} shows the result for Si, Ge, Si$_{0.99}$Ge$_{0.01}$ and Si$_{0.2}$Ge$_{0.8}$ against the data of Ref.~\cite{Wilson2014}, and the apparent interface conductance of Al/Si$_{0.99}$Ge$_{0.01}$ against Ref.~\cite{Wilson2013}. For Ge and Si$_{0.2}$Ge$_{0.8}$ the adiabatic condition reproduces $\Lambda_A(f)$ in level and slope ($54\to42$ and $8.9\to4.9$~W\,m$^{-1}$K$^{-1}$ between 1.1 and 17.6~MHz against $53\to42$ and $7.8\to4.7$ measured), Si$_{0.99}$Ge$_{0.01}$ falls between the two conditions, and the frequency dependence is a weak power law rather than a Lorentzian, as in the data. For Si both conditions overpredict the through-plane suppression with the log-normal spectrum of Sec.~\ref{sec:spectra} ($145\to117$ against $137\to128\pm10$, thermalizing condition). The result is sensitive to the fraction of $\kappa$ carried by modes with $\Lambda>1~\mu$m: the Born--von K\'arm\'an/Callaway spectrum with natural-isotope scattering (24\% instead of 40\% above $1~\mu$m, Sec.~\ref{sec:spectra}) gives $144\to125$~W\,m$^{-1}$K$^{-1}$ with the thermalizing condition [dash-dotted in Fig.~\ref{fig:S2}(a); $139\to114$ with modes below 3~THz adiabatic], within the error bars of all six points but with twice the measured slope ($13\%$ against $7\pm8\%$). Monte Carlo solutions of the single-pulse problem likewise predict a cross-plane suppression that the experiment does not show~\cite{Ding2014}, whereas the analytic RTA solution of Koh \emph{et al.}~\cite{Koh2014} for a semi-infinite solid under periodic heating, without transducer or interface, finds the out-of-phase apparent conductivity of pure crystals nearly independent of $f$, the reduced amplitude and the reduced phase lag of the heat flux compensating each other. The present solver, which fits $(\Lambda_A,G_A)$ to the full in-phase/out-of-phase ratio with the transducer and $G_{\mathrm{int}}$ in place, recovers this compensation only partially, the more so the smaller the long-MFP fraction. The two Si spectra thus pull in opposite directions: the in-plane quantities of Fig.~\ref{fig:S1} favour the long-MFP-rich spectrum, the through-plane response the Born--von K\'arm\'an one, and no single RTA spectrum with a sector-averaged interface condition reproduces both. We therefore do not claim the through-plane response of Si, which Ref.~\cite{Hua2019} reproduces with the measured spectral transmissivity of the Al/Si interface. \emph{Nonequilibrium interfacial resistance: definition and values.} We define $R_{\mathrm{neq}}\equiv\Delta T_{\mathrm{neq}}/q''$, where $q''$ is the (periodic) heat flux per unit area through the interface and $\Delta T_{\mathrm{neq}}$ is the \emph{additional} temperature drop, at the modulation frequency, that the local two-parameter model must attribute to the interface in order to reproduce the signal generated by the nonlocal substrate: operationally, $R_{\mathrm{neq}}(f)=1/G_A(f)-1/G_{\mathrm{int}}$, with $G_{\mathrm{int}}$ the local conductance put into the forward model. $\Delta T_{\mathrm{neq}}$ is thus the difference between the interface temperature the transducer sees and the one a Fourier substrate of conductivity $\Lambda_A$ would give for the same flux; it is a property of the substrate's nonequilibrium layer, not of the physical interface, and it is frequency dependent because $d_p(f)$ sets how much of that layer the measurement weights. With $G_{\mathrm{int}}=250$~MW\,m$^{-2}$K$^{-1}$ and a thermalizing interface the fitted $G_A$ of Al/Si$_{0.99}$Ge$_{0.01}$ rises from 96 at 1.1~MHz to 144~MW\,m$^{-2}$K$^{-1}$ at 17.6~MHz, i.e.\ $R_{\mathrm{neq}}=6.5\to3.0\times10^{-9}$~m$^2$K\,W$^{-1}$; for the other materials at 1.1~MHz, $R_{\mathrm{neq}}=3.0$ (Si), $2.5$ (Ge) and $12\times10^{-9}$~m$^2$K\,W$^{-1}$ (Si$_{0.2}$Ge$_{0.8}$), decreasing by a factor 1.5--2.2 at 17.6~MHz, and larger by 30--60\% with the adiabatic condition. For comparison, measured Kapitza resistances $R_K=1/G$ are $3$--$5\times10^{-9}$~m$^2$K\,W$^{-1}$ for Al/Si ($G=350$, $223$ and $250$~MW\,m$^{-2}$K$^{-1}$ in Refs.~\cite{Minnich2011,Hua2019,Wilson2014}; the value $G=110$~MW\,m$^{-2}$K$^{-1}$ assumed in the simulations of Ref.~\cite{Ding2014} would correspond to $9\times10^{-9}$) and $5$--$7\times10^{-9}$~m$^2$K\,W$^{-1}$ for Al/Si$_{0.99}$Ge$_{0.01}$~\cite{Wilson2013,Wilson2014}: $R_{\mathrm{neq}}$ is of the same order as $R_K$, comparable to it in the alloys and smaller in Si and Ge. This compatibility of magnitude supports the plausibility of the mechanism; it does not by itself establish its microscopic origin, nor does it imply that the nonequilibrium layer accounts for the whole measured resistance (the comparison with the absolute level of $G_A$ below shows that it does not). The excess resistance is the ballistic injection resistance of the nonlocal modes, the interfacial nonequilibrium resistance of the Letter, and lies in the range $(5$--$9)\times10^{-9}$~m$^2$K\,W$^{-1}$ inferred in Ref.~\cite{Wilson2013} from the two-channel analysis; relative to the measured $G_A$ itself ($\approx155\to190$~MW\,m$^{-2}$K$^{-1}$, which with the fitted $G_2=320$ corresponds to $1/G_A-1/G_2\approx3.5\to2\times10^{-9}$~m$^2$K\,W$^{-1}$) the thermalizing condition overestimates the injection resistance by about a factor of two at 1~MHz, i.e.\ the real Al/SiGe interface injects the long-MFP modes only partially, the same conclusion reached in Ref.~\cite{Wilson2013} ($G_1\approx30$~MW\,m$^{-2}$K$^{-1}$), while the frequency increment of $G_A$ ($+48$ computed, $+35$~MW\,m$^{-2}$K$^{-1}$ measured between 1 and 18~MHz) is reproduced. The calculation thus separates two contributions: the frequency dependence of the \emph{apparent} conductance is a substrate-transport effect absorbed by the local model, whereas its absolute level is set by the genuinely interfacial transmissivity of the low-frequency modes.

% ======================================================================
\suppsection{Spectral classification of transport regimes}{sec:spectral}

The bath spectral density fixes the memory kernel via $\bm{\Phi}(t)=(2/\pi)\int_0^\infty[\mathbf{S}(\omega)/\omega]\cos(\omega t)\,d\omega$. A power law $\mathbf{S}(\omega)\propto\omega^{s}$ gives $\bm{\Phi}(t)\propto t^{-s}$ and, in the Laplace domain, $\widetilde{\bm{\Phi}}(z)\sim\mathbf{C}_s z^{s-1}$ ($z\to0$)~\cite{Morgado2002,Lapas2007,Vainstein2006}. The final-value theorem applied to $\widetilde{\mathbf{R}}(z)=[z\mathbf{1}+\widetilde{\bm{\Phi}}(z)]^{-1}$ (homogeneous relaxation, $\mathbf{k}=0$, where $\bm{\Omega}=\mathbf{0}$) gives $\lim_{t\to\infty}\mathbf{R}(t)=\lim_{z\to0}[\mathbf{1}+\mathbf{C}_s z^{s-2}]^{-1}$, classifying the dynamics:
\begin{enumerate}
\item \emph{Ohmic, $s=1$ (normal diffusion).} $\widetilde{\bm{\Phi}}(0)=\bm{\gamma}^\ast$ finite; $\mathbf{R}(\infty)=\mathbf{0}$ (exponential relaxation), recovering Fourier's law and the generalized Einstein relation $\mathbf{L}_{\mathrm{Onsager}}=(k_BT)^{-1}[\bm{\gamma}^\ast]^{-1}\mathbf{C}_0$.
\item \emph{Sub-Ohmic, $0<s<1$ (subdiffusion).} $\widetilde{\bm{\Phi}}(z)\sim z^{s-1}$ diverges; ergodic but algebraically slow relaxation, $\alpha=s<1$, where $\alpha$ is known as diffusion exponent.
\item \emph{Super-Ohmic, $1<s<2$ (superdiffusion).} Persistent correlations with $\alpha>1$; the system still relaxes, $\mathbf{R}(\infty)=\mathbf{0}$.
\item \emph{Ballistic, $s=2$.} $\mathbf{R}(\infty)=[\mathbf{1}+\mathbf{C}_s]^{-1}\neq\mathbf{0}$: the autocorrelation never vanishes and the hydrodynamic description breaks down~\cite{Lapas2008}.
\end{enumerate}
Beyond $s>2$, concentration-dependent diffusivity $D(C)\sim C^{-\gamma}$ produces hyperballistic spreading that requires a nonlinear treatment~\cite{Flekkoy2021}. The Cattaneo--Vernotte equation~\cite{Cattaneo1948,Vernotte1958} follows from the Ohmic (Markovian) closure itself: a constant memory $\widetilde{\bm{\Phi}}(z)\approx\tau_q^{-1}\mathbf{1}$ gives $\widetilde{\mathbf{R}}(z)$ a single inertial pole at $z=-\tau_q^{-1}$, hence $\widetilde{\mathbf{L}}(z)=\mathbf{L}_0/(1+z\tau_q)$ and, in the time domain, $\tau_q\,\partial_t\mathbf{J}_q+\mathbf{J}_q=-\kappa\nabla T$; Fourier's law is its limit $\omega\tau_q\ll1$. This temporal-memory regime ($\omega\tau_q\gtrsim1$) is that of second sound, including transient gratings in the hydrodynamic window~\cite{Hua2020,Zeng2026}; room-temperature transient gratings~\cite{Maznev2011}, like TDTR, operate at times far longer than $\tau_q$ and probe spatial nonlocality, the quasistatic regime of the analysis below.

% ======================================================================
\suppsection{Anisotropy, wavevector nonlocality, and the interface}{sec:aniso}

\emph{Anisotropic conductivity.}---Fourier transforming Eq.~\eqref{eq:Lqq-quantum} in space and time and dividing by $T_0^2$ gives the generalized conductivity tensor
$\kappa_{\mathrm{eff}}^{ij}(\mathbf{k},\omega)=\widetilde L_{qq}^{ij}(\mathbf{k},\omega)/T_0^2$, in which the $T_0^2$ of Eq.~\eqref{eq:Lqq-quantum} cancels; for extended modes the sum is per unit volume and the $\mathbf{k}$ dependence sits in the propagator (Sec.~\ref{sec:rta}).
Because $\bm{\Psi}_\mu$ carries the directional group velocity $\mathbf{v}_\mu$, the diagonal blocks are weighted by the corresponding velocity projections,
\begin{equation}
\kappa_{\mathrm{eff}}^{ii}(\mathbf{k},\omega)=\frac{1}{V}\sum_{\mu,\nu}\widetilde\Psi_\mu^{i}(\mathbf{k})\,\widetilde\Psi_\nu^{i}(-\mathbf{k})\,\widetilde R_{\mu\nu}(\mathbf{k},\omega)\,c_{\mathrm{ph}}(\omega_\nu,T),
\label{eq:kappa-tensor}
\end{equation}
and are in general unequal, $\kappa_{\mathrm{eff}}^{zz}\neq\kappa_{\mathrm{eff}}^{rr}$, even for a crystal whose bulk conductivity is isotropic.

In a TDTR experiment, the measured signal is, strictly, sensitive to a distribution of in-plane spatial frequencies (Hankel-space integral)~\cite{Feser2012}. It is nevertheless convenient to summarize the dominant gradients by effective wavevectors: $k_r\sim c/w_0$ (spot size, $c\approx2.4$ for the beam-offset width at $w_0\approx1~\mu$m) and a through-plane wavevector $k_z\sim d_p^{-1}(f)$ associated with the oscillatory penetration depth. To avoid circularity when the apparent conductivity depends on $f$, we define
\begin{equation}
d_p(f)\equiv \sqrt{\frac{\kappa_{\mathrm{ref}}}{\pi C f}},
\qquad
k_z\sim d_p^{-1}(f),
\label{eq:dp-kz}
\end{equation}
where $\kappa_{\mathrm{ref}}$ is a fixed diffusive reference conductivity (taken as $\kappa_{\mathrm{ref}}=\kappa_{\mathrm{bulk}}$ in the numerical estimates below). Thus,
\begin{equation}
k_r\simeq c/w_0,\qquad c\simeq2.3\text{--}3.6\ \text{for}\ w_0=1\text{--}10~\mu\mathrm{m}\ \text{(beam-offset width, Sec.~\ref{sec:spectra})}.
\label{eq:kr}
\end{equation}
The measured through- and in-plane conductivities,
\begin{equation}
\Lambda_z=\kappa_{\mathrm{eff}}^{zz}(k_z),\qquad \Lambda_r=\kappa_{\mathrm{eff}}^{rr}(k_r),
\label{eq:Lz-Lr}
\end{equation}
differ, with the more strongly sampled direction the more suppressed. For Si at $w_0\approx1\,\mu$m, $f=9.8$~MHz, $\kappa_{\mathrm{ref}}\approx142~\mathrm{W\,m^{-1}\,K^{-1}}$ and $C\approx1.6\times10^{6}~\mathrm{J\,m^{-3}\,K^{-1}}$, one finds $d_p\approx1.7~\mu$m, hence $k_z\approx0.6~\mu\mathrm{m}^{-1}<k_r\approx2.4~\mu\mathrm{m}^{-1}$ and $\Lambda_r<\Lambda_z$, consistent with the beam-offset values $\Lambda_z\approx140$, $\Lambda_r\approx80~\mathrm{W\,m^{-1}\,K^{-1}}$~\cite{Wilson2014,Feser2012}.

\emph{Spatial-nonlocality reduction.}---A minimal, one-length parametrization consistent with the long-wavelength expansion in Sec.~SIV.A is
\begin{equation}
\kappa_{\mathrm{eff}}(k)=\frac{\kappa_{\mathrm{bulk}}}{1+(k\ell)^2},
\label{eq:kspace-SM}
\end{equation}
with $\ell$ the nonlocality range (an effective MFP-like length).
Evaluating at $k=k_z\sim d_p^{-1}(f)$ with Eq.~\eqref{eq:dp-kz} yields the frequency-resolved apparent conductivity
\begin{equation}
\Lambda_A(f)=\frac{\kappa_{\mathrm{bulk}}}{1+f/f^\ast},
\qquad
f^\ast=\frac{\kappa_{\mathrm{ref}}}{\pi C\,\ell^2}.
\label{eq:LambdaA-SM}
\end{equation}
This is the two-parameter model ($\kappa_{\mathrm{bulk}},f^\ast$, with $\ell=d_p(f^\ast)$ derived) fitted in the Letter against the frequency-dependent TDTR apparent conductivities of Wilson and Cahill~\cite{Wilson2014} (for their concentric-beam data at $w_0=10~\mu$m). The crossover $f^\ast$ scales as $\ell^{-2}$: high-conductivity materials (pure Si, Si:B) keep $f^\ast$ above the band and stay flat although their $\ell$ is not shorter than that of Ge, while Ge and the SiGe alloys lower $f^\ast$ into the measurement window and roll off.

Using digitized data of Ref.~\cite{Wilson2014}, representative fits yield $\ell=0.41\pm0.09$~($\mathrm{Ge}$), $0.34\pm0.07$~(Si$_{0.99}$Ge$_{0.01}$) and $0.25\pm0.04~\mu$m~(Si$_{0.2}$Ge$_{0.8}$), with $\chi^2_\nu<1$; Si and Si:B are flat and leave $\ell$ unresolved (Sec.~\ref{sec:fitproc} and Table~\ref{tab:fits} give the fitting procedure, the fixed inputs, the statistical and systematic uncertainties, and the meaning of the quoted range). Two trends follow from Eq.~\eqref{eq:Lqq-quantum}: mass-disorder scattering ($\propto\omega^4$) removes high-frequency modes and increases the fraction of $\kappa$ carried by long-MFP modes~\cite{Koh2007}, so the alloys roll off within the band while Si does not (the fitted $\ell$ measures the MFP window sampled by $d_p(f)$, not the mean MFP); and the lower alloy conductivity shrinks $d_p$ and increases the probed $k_z\sim d_p^{-1}$. Because $\ell$ is a length, it can be compared with the conductivity accumulation function $\alpha(L)$~\cite{Wilson2014}: the sub-micron values lie within the MFP range that dominates $\alpha(L)$ in these crystals.

\suppsubsection{Fitting procedure, fixed inputs and uncertainties}{sec:fitproc}
\emph{Data.} The six $(f,\Lambda_A)$ points per material were digitized from Fig.~3a of Ref.~\cite{Wilson2014} ($w_0=10~\mu$m), with the error bars shown there; a digitization error of $\sim3\%$ of $\Lambda_A$ is small compared with those bars.

\emph{Free and fixed parameters.} Equation~\eqref{eq:LambdaA-SM} is fitted by weighted least squares with \emph{two} free parameters per material, $\kappa_{\mathrm{bulk}}$ and $f^\ast$; the nonlocality length is derived, $\ell=[\kappa_{\mathrm{bulk}}/(\pi Cf^\ast)]^{1/2}$, with the volumetric heat capacity held fixed at $C=1.63$ (Si, Si:B, Si$_{0.99}$Ge$_{0.01}$), $1.67$ (Ge) and $1.66$~MJ\,m$^{-3}$K$^{-1}$ (Si$_{0.2}$Ge$_{0.8}$), standard room-temperature values (Ref.~\cite{Wilson2013} uses $1.65$ for Si$_{0.99}$Ge$_{0.01}$, a 1\% difference). Since $\kappa_{\mathrm{ref}}=\kappa_{\mathrm{bulk}}$ in $d_p$, the fit determines $f^\ast$ directly and $\ell$ is a reparametrization, $\ell=d_p(f^\ast)$.

\emph{Statistical uncertainties.} The quoted $\pm$ are one-standard-deviation values from the covariance of the weighted fit, treating the digitized error bars as absolute $1\sigma$ errors; for $\ell$ they are propagated with the full covariance ($\kappa_{\mathrm{bulk}}$ and $f^\ast$ are correlated, $r\approx-0.8$) and agree within $0.01~\mu$m with the $\Delta\chi^2=1$ intervals of the profile likelihood in $\ell$ ($0.33$--$0.50$, $0.27$--$0.42$ and $0.21$--$0.28~\mu$m for Ge, Si$_{0.99}$Ge$_{0.01}$ and Si$_{0.2}$Ge$_{0.8}$). $\chi^2_\nu=0.05$--$0.5$ for all materials, so these bars are conservative and the statistical uncertainties are, if anything, overestimated. For Si and Si:B the roll-off is within the error bars, $f^\ast$ is unresolved at the $1\sigma$ level ($177\pm180$ and $362\pm694$~MHz), and $\ell$ is an upper bound: $\ell_{\mathrm{Si}}\lesssim0.6~\mu$m ($1\sigma$; the profile likelihood gives $0.58~\mu$m at $1\sigma$ and $0.75~\mu$m at $2\sigma$).

\emph{Meaning of the range $0.25$--$0.4~\mu$m.} It is the spread of the three resolved values across materials (Ge, Si$_{0.99}$Ge$_{0.01}$, Si$_{0.2}$Ge$_{0.8}$), not a confidence interval of a single quantity.

\emph{Systematic uncertainties.} Two choices affect $\ell$ more than the statistics. (a) The functional form: evaluating the Lorentzian at the complex wavevector of the thermal wave, $k_z^2=-2i/d_p^2$, and taking the modulus, $\Lambda_A=\kappa_{\mathrm{bulk}}/[1+4(f/f^\ast)^2]^{1/2}$, fits the six points equally well and returns $\ell=0.47$, $0.38$ and $0.22~\mu$m ($\pm15\%$ relative to the real-$k$ form). (b) The treatment of $\kappa_{\mathrm{bulk}}$: fixing it at literature values ($60$, $42$ and $9$~W\,m$^{-1}$K$^{-1}$) instead of fitting it transfers the suppression already present at the lowest frequency to $f^\ast$ and gives $\ell=0.55$, $0.56$ and $0.34~\mu$m. The fitted $\kappa_{\mathrm{bulk}}$ ($55$, $36$ and $7.8$) lie $8$--$15\%$ below the literature values, which is the signature of nonlocal suppression at $k_z=d_p^{-1}(1~\mathrm{MHz})$ already noted in the Letter. Accordingly, $\ell$ should be read as $\ell_{\mathrm{fit}}\pm0.15~\mu$m (systematic), the systematic part dominating. \emph{Joint test with a shared $\ell$.} A natural test of the one-length form is whether a single $\ell$ for Si at 300~K describes both the beam-offset anisotropy and the high-frequency through-plane data. With $k_r=w_0^{-1}$ it does not: $\Lambda_r=80$~W\,m$^{-1}$K$^{-1}$ at $w_0=1~\mu$m requires $\ell\approx0.88~\mu$m, which gives $\Lambda_z=112$ and $92$~W\,m$^{-1}$K$^{-1}$ at 9.8 and 20~MHz with $k_z=d_p^{-1}$, against $\approx140$ and $128\pm10$ measured. With the effective in-plane wavevector obtained from the forward model, $k_r\simeq2.4/w_0$ (Sec.~\ref{sec:spectra}), the same $\Lambda_r$ requires $\ell=0.88/2.4\approx0.37~\mu$m, and this $\ell$ gives $\Lambda_z=136$ and $130$~W\,m$^{-1}$K$^{-1}$ at 9.8 and 20~MHz, within the error bars and consistent with $\ell_{\mathrm{Si}}\lesssim0.6~\mu$m of Table~\ref{tab:fits}. The one-length form therefore passes the joint test once the in-plane wavevector is taken from the forward model, but the test has little discriminating power: $\Lambda_z(f)$ of Si is flat, so any $\ell\lesssim0.6~\mu$m is admissible, and the through-plane response is governed by the interfacial condition rather than by $\kappa(d_p^{-1})$ (Sec.~\ref{sec:bvp}). The limitation of the Lorentzian is one of representation, not of this estimate: it cannot represent a kernel that decays almost logarithmically in $k$ over the TDTR window [Fig.~\ref{fig:S1}(a)], and its $\ell$ is an effective range rather than $\ell_L=(\tfrac35\langle\Lambda^2\rangle_\kappa)^{1/2}$. Section~\ref{sec:bvp} and Fig.~\ref{fig:S1} replace it by the full kernel in the full TDTR model. The fits of Fig.~2 are therefore to be read as \emph{compatibility} of the data with a one-length nonlocal description, not as a confirmation of the theory; the confirmation, to the extent provided, is the parameter-free comparison of Figs.~\ref{fig:S1}--\ref{fig:S2}, which uses no fitted $\ell$.

\begin{table}[t]
\caption{Fits of Eq.~\eqref{eq:LambdaA-SM} to the data of Ref.~\cite{Wilson2014}: free parameters $\kappa_{\mathrm{bulk}}$ and $f^\ast$ ($1\sigma$ from the covariance), derived $\ell$ ($1\sigma$ with the full covariance), reduced $\chi^2$, and the two systematic variants described in Sec.~\ref{sec:fitproc}.}
\label{tab:fits}
\begin{ruledtabular}
\begin{tabular}{lccccccc}
 & $C$ (MJ\,m$^{-3}$K$^{-1}$) & $\kappa_{\mathrm{bulk}}$ (W\,m$^{-1}$K$^{-1}$) & $f^\ast$ (MHz) & $\ell$ ($\mu$m) & $\chi^2_\nu$ & $\ell$, complex-$k$ form & $\ell$, $\kappa_{\mathrm{bulk}}$ fixed \\ \hline
Si & 1.63 & $142\pm7$ & $177\pm180$ & $0.40\pm0.21$ & 0.08 & 0.59 & 0.39 (142) \\
Si:B & 1.63 & $113\pm5$ & $362\pm694$ & $0.25\pm0.24$ & 0.46 & 0.45 & 0.11 (110) \\
Ge & 1.67 & $55\pm2$ & $61\pm24$ & $0.41\pm0.09$ & 0.22 & 0.47 & 0.55 (60) \\
Si$_{0.99}$Ge$_{0.01}$ & 1.63 & $36\pm2$ & $60\pm24$ & $0.34\pm0.07$ & 0.05 & 0.38 & 0.56 (42) \\
Si$_{0.2}$Ge$_{0.8}$ & 1.66 & $7.8\pm0.4$ & $25\pm6$ & $0.25\pm0.04$ & 0.43 & 0.22 & 0.34 (9) \\
\end{tabular}
\end{ruledtabular}
\end{table}

The same frequency window can often be fit by a temporal (Cattaneo) form $\kappa_{\mathrm{bulk}}/[1+(2\pi f\tau)^2]$; the spatial interpretation is preferred not because the two fits are necessarily distinguishable over a narrow band, but because TDTR typically operates at $1/f\gg\tau_{\mathrm{carrier}}$, so temporal memory is negligible and $\ell$ is the physically meaningful parameter.

\emph{Interfacial nonequilibrium resistance.}---Near a metal/sample boundary the reflected and transmitted modes truncate the overlap $\Psi_\mu^{i}(\mathbf{r})\Psi_\nu^{j}(\mathbf{r}')$ within a layer of thickness $\sim\ell$, and the off-diagonal correlations $R_{\mu\nu}$ ($\mu\neq\nu$) between short- and long-MFP sectors fail to relax to a single temperature there. This is the mechanism behind the interfacial nonequilibrium resistance inferred from the frequency dependence of $G_A$ in SiGe alloys~\cite{Wilson2014,Wilson2013}; its quantitative form, the two-sector projection and the computed $R_{\mathrm{neq}}$ are given in Secs.~\ref{sec:bvp} and \ref{sec:twosector}.

% ======================================================================
\suppsection{Two-sector projection and the two-channel model}{sec:twosector}
Partition the slow variables into a long-MFP sector 1 (modes below 3~THz in the alloys~\cite{Wilson2013}) and a diffusive reservoir 2, and project Eq.~\eqref{eq:GLE} onto the sector energies $E_1,E_2$. The dissipative block $\Phi_{12}=\Phi_{21}\equiv g$ is the intersector relaxation rate (energy transfer per unit volume and kelvin) and the intrasector streaming yields the sector conductivities $\Lambda_{1,2}$. The resulting equations,
\begin{equation}
C_1\partial_tT_1=\Lambda_1\nabla^2T_1-g(T_1-T_2),\qquad C_2\partial_tT_2=\Lambda_2\nabla^2T_2+g(T_1-T_2),
\end{equation}
are Eqs.~(1)--(2) of Ref.~\cite{Wilson2013}, with nonequilibrium length $d=[g(\Lambda_1^{-1}+\Lambda_2^{-1})]^{-1/2}$ and relaxation time $\Gamma=[g(C_1^{-1}+C_2^{-1})]^{-1}$. The ballistic/diffusive variant of the same reference replaces sector 1 by a stream with exponential kernel $e^{-|z-z'|/\ell_1}$ (a Lorentzian in $k$) and gives the same decay length with $d=\sqrt{\Lambda_2/(\Lambda_1+\Lambda_2)}\,\ell_1$ [their Eq.~(21)]. With $\Lambda_1=30$, $\Lambda_2=12$~W\,m$^{-1}$K$^{-1}$ and $d=150$~nm~\cite{Wilson2013} this yields $\ell_1=0.28~\mu$m for the range of the ballistic sector. Because only sector 1 is nonlocal, the total conductivity is $\Lambda_2+\Lambda_1/(1+k^2\ell_1^2)$, whose one-length form, Eq.~\eqref{eq:kspace-SM}, has $\ell=\ell_1[\Lambda_1/(\Lambda_1+\Lambda_2)]^{1/2}\approx0.24~\mu$m; both values are consistent with $\ell_{\mathrm{Si_{0.99}Ge_{0.01}}}=0.34\pm0.07$ (statistical) $\pm0.15~\mu$m (systematic) of the Letter. In the adiabatic limit for sector 1, $G_{\mathrm{NE}}=\Lambda_2(\Lambda_1+\Lambda_2)/(\Lambda_1d)=112$~MW\,m$^{-2}$K$^{-1}$ with these parameters [Eq.~(10) of Ref.~\cite{Wilson2013}], while the same reference quotes $G_{\mathrm{NE}}\approx200$~MW\,m$^{-2}$K$^{-1}$ from its fit of $\Lambda_A$ and $G_A$; i.e.\ $R_{\mathrm{NE}}=(5$--$9)\times10^{-9}$~m$^2$K\,W$^{-1}$, in agreement with the direct calculation of Sec.~\ref{sec:bvp}.

% ======================================================================
\suppsection{Mean-free-path spectra and TDTR forward model}{sec:spectra}
\emph{Si.}---The conductivity accumulation $\alpha(\Lambda)$ at 300~K is represented by a log-normal distribution in $\ln\Lambda$ anchored to three published values, from first principles~\cite{Esfarjani2011} and from MFP spectroscopy~\cite{Regner2013}: $\alpha(0.1~\mu\mathrm{m})=0.10$, $\alpha(1~\mu\mathrm{m})=0.60$ (40\% of $\kappa$ from $\Lambda>1~\mu$m) and $\alpha(8~\mu\mathrm{m})=0.95$, giving a median of $0.68~\mu$m and $\sigma_{\ln}=1.5$; $\kappa_{\mathrm{bulk}}=148$~W\,m$^{-1}$K$^{-1}$, $C=1.63$~MJ\,m$^{-3}$K$^{-1}$, and $v=6000$~m/s for the bin velocities. A Born--von K\'arm\'an/Callaway model with natural-isotope scattering (Tamura strength, $B$ fixed by $\kappa=148$~W\,m$^{-1}$K$^{-1}$), which gives $\alpha(1~\mu\mathrm{m})=0.76$, is used to bracket the uncertainty: it yields $121\to128$ and $70\to92$~W\,m$^{-1}$K$^{-1}$ for the two in-plane Si quantities quoted in the Letter, and $144\to125$~W\,m$^{-1}$K$^{-1}$ for $\Lambda_A(f)$ between 1.1 and 17.6~MHz [Fig.~\ref{fig:S2}(a), dash-dotted; Sec.~\ref{sec:bvp}].

\emph{Ge and Si$_{1-x}$Ge$_x$.}---Isotropic Born--von K\'arm\'an dispersion (LA and two TA branches; $v_{\mathrm{LA,TA}}=4920,3540$~m/s for Ge, $8430,5840$~m/s for Si, linearly interpolated for the alloys), scattering rate $\tau^{-1}=A\omega^4+BT\omega^2e^{-C_U/T}$ with Tamura mass-disorder strength~\cite{Tamura1983} $A=V_0\Gamma/4\pi\bar v^3$, $\Gamma=x(1-x)(\Delta M/\bar M)^2+\Gamma_{\mathrm{iso}}$ ($\Gamma=0.024$ for $x=0.01$ and $0.079$ for $x=0.8$; $g_{\mathrm{iso}}=5.9\times10^{-4}$ for Ge), $C_U=137$~K (Si) and 75~K (Ge), and $B$ fixed by $\kappa_{\mathrm{Ge}}=60$~W\,m$^{-1}$K$^{-1}$ ($B=3.1\times10^{-19}$~s/K) and interpolated for the alloys; modes below 0.5~THz are removed (Akhiezer damping, as in Ref.~\cite{Wilson2013}). The resulting intrinsic conductivities, 56.7~W\,m$^{-1}$K$^{-1}$ (Si$_{0.99}$Ge$_{0.01}$) and 13.8~W\,m$^{-1}$K$^{-1}$ (Si$_{0.2}$Ge$_{0.8}$), are outputs, not inputs. The fraction of $\kappa$ carried by modes with $\Lambda>1~\mu$m is $16\%$ (Ge), $47\%$ (Si$_{0.99}$Ge$_{0.01}$) and $33\%$ (Si$_{0.2}$Ge$_{0.8}$), against $24\%$ for the Born--von K\'arm\'an Si spectrum above: mass disorder raises the long-MFP fraction. Modes are grouped into 26 logarithmic MFP bins with $\kappa$-weighted mean $\Lambda_b$ and $C$-weighted mean $v_b$; the bin heat capacity is set so that $\kappa_b=C_bv_b^2\tau_b/3$. Replacing these semi-empirical spectra by tabulated first-principles lifetimes $\tau_s(\mathbf{q})$ and velocities $\nabla_{\mathbf{q}}\omega_s(\mathbf{q})$ requires no extension of the model: the relaxation-time closure of Sec.~\ref{sec:rta} takes the mode set $\{C_\mu,\mathbf{v}_\mu,\tau_\mu\}$ as input, and the scripts accept such arrays directly. It is left for future work only because the required data are not tabulated in the literature in a directly usable form (the almaBTE database~\cite{Carrete2017}, for instance, provides force constants from which they must be computed).

\emph{Forward model.}---Cahill's Hankel-space multilayer solution with Gaussian pump and probe~\cite{Cahill2004,Feser2012}: Al transducer 70~nm with $G=110$~MW\,m$^{-2}$K$^{-1}$ for the spot-size and beam-offset calculations (parameters of Ref.~\cite{Ding2014}), 80~nm with $G_{\mathrm{int}}=250$~MW\,m$^{-2}$K$^{-1}$ and $w_0=w_1=10~\mu$m for the frequency study; repetition rate 80~MHz, $|m|\le150$ harmonics, delay window 0.1--4~ns. For the spot-size and beam-offset calculations the substrate in-plane conductivity is $\kappa_L(2\pi k)$, $k$ being the Hankel variable, and the through-plane conductivity is either $\kappa_{\mathrm{bulk}}$ or $\kappa_T(2\pi k)$; the in-plane conductivity is extracted from the full width at half maximum of the out-of-phase signal versus beam offset at 100~ps delay with the through-plane conductivity held at the bulk value~\cite{Feser2012,Wilson2014}. Inverting $\kappa_L(k_{\mathrm{eff}})=\Lambda_r$ gives, for the beam-offset width, $k_{\mathrm{eff}}w_0=2.3$, $2.4$, $2.8$, $3.4$ and $3.6$ at $w_0=1$, $1.5$, $2.4$, $5$ and $10$--$12~\mu$m ($\Lambda_r=70$, $82$, $92$, $109$ and $124$--$127$~W\,m$^{-1}$K$^{-1}$): $c\simeq2.4$ holds at $w_0\approx1$--$1.5~\mu$m, the case discussed in the Letter, and $c$ grows slowly with $w_0$ because the signal averages $\kappa_L$ over a distribution of $k$, whose inversion through the nonlinear $\kappa_L(k)$ depends on the local curvature of the kernel. For the concentric fit $k_{\mathrm{eff}}D\simeq1.5$ at $D=3.7~\mu$m and varies between $0.6$ and $1.8$ for $D=2$--$30~\mu$m. The effective wavevector is thus a property of the geometry and of the kernel, and different experiments are to be compared through the forward model rather than through a fixed $c$ [Sec.~\ref{sec:pred}, prediction (ii)]. Python scripts reproducing Figs.~\ref{fig:S1}--\ref{fig:S2} are provided as Supplemental files.

\suppsubsection{Scope: ingredients, approximations and limitations}{sec:scope}
To make explicit what the theory resolves, what it replaces by effective quantities, and what it leaves out, we separate three levels.

\emph{Formal level [Eqs.~(2)--(7) of the Letter].} Phonon dispersion enters through $\omega_\mu$ (in $c_{\mathrm{ph}}$) and the group velocities $\mathbf{v}_\mu$ (in $\bm{\Psi}_\mu$); relaxation enters through the propagator $R_{\mu\nu}(\mathbf{k},t)$, which is left unspecified. Nothing is averaged at this level, but nothing is computed either: the kernel is an exact rewriting of the Green--Kubo correlation in the chosen basis, and its content is fixed only by a closure of $\bm{\Phi}$. Positivity of $\sigma$ (Secs.~\ref{sec:hydro} and \ref{sec:modal}) is the only statement that does not depend on the closure, provided the closure respects the fluctuation--dissipation theorem.

\emph{One-length level [Eqs.~(10)--(11) of the Letter, Fig.~2].} Mode details are replaced by a single effective length. Section~\ref{sec:rta} shows what this length is: the $[0/1]$ Pad\'e closure of $\kappa_L(k)$, with $\ell^2=\tfrac35\langle\Lambda^2\rangle_\kappa$ formally, and in practice, because the Pad\'e form is controlled only for $k\Lambda_{\max}\lesssim1$, an effective range that a Lorentzian returns when fitted over the measured decade ($\ell_{\mathrm{eff}}\approx0.6~\mu$m for the Si kernel between $k=0.3$ and $1.2~\mu$m$^{-1}$). The fits of Fig.~2 therefore encode the mode-resolved physics only through two numbers per material, and the discussion in the Letter of trends with alloying is to be read at that level.

\emph{Mode-resolved level [Secs.~\ref{sec:rta}--\ref{sec:spectra}, Figs.~S1--S2].} The relaxation-time closure resolves $(C_\mu,v_\mu,\tau_\mu)$ explicitly (26 MFP bins, 8 directions). Included: dispersion and group velocities (isotropic Born--von K\'arm\'an for Ge and the alloys; an accumulation function for Si anchored to published first-principles and MFP-spectroscopy values); Bose statistics; isotope, mass-disorder and umklapp scattering; the longitudinal/transverse structure of the kernel; the modal boundary condition at the interface in two limiting forms; the complete multilayer, pulse-accumulated TDTR signal and its analysis with the local model. Replaced by effective or external quantities: the sector-averaged interface condition (thermalizing, or adiabatic below 3~THz, for Si for $\Lambda>0.7~\mu$m, rather than the spectral transmissivity of Ref.~\cite{Hua2019}); $G_{\mathrm{int}}$; the bin velocity of the Si spectrum ($v=6000$~m/s). Omitted: intermode (off-diagonal) relaxation beyond the RTA, hence the normal/umklapp distinction and phonon hydrodynamics below $\sim100$~K; anisotropy of the dispersion (cubic averaging); optical phonons except through the heat capacity; nondiffusive transport in the Al transducer and electron--phonon coupling; the three-dimensional structure of the through-plane problem (treated in 1D, valid for $w_0=10~\mu$m); temperature dependence (300~K only). None of the quantities in the comparisons of Figs.~S1--S2 is adjusted to the TDTR data.

% ======================================================================
\suppsection{Relation to previous theoretical treatments}{sec:relation}
Table~\ref{tab:relation} places the present framework next to the treatments that share its goal. Three points summarize the comparison. (i) \emph{Starting point.} All previous treatments start from a kinetic equation with a given collision operator: the phonon BTE in the relaxation-time approximation (RTA)~\cite{Mahan1988,Koh2014,Hua2014,Hua2019,Ding2014,YangDames2015} or with the full collision matrix~\cite{Hua2020}, its Guyer--Krumhansl reduction~\cite{Beardo2022,Xiang2022}, a two-channel diffusive model~\cite{Wilson2013}, or, closest to the present work, a spatiotemporal Green--Kubo kernel defined by Zwanzig/Robertson projection and realized for phonons through the linearized BTE~\cite{Zeng2026}. The present work starts from entropy production in phase space (MNET) and the Mori--Kubo--Zwanzig identity for the Liouville dynamics; the kinetic kernels are recovered as closures of the memory function $\bm{\Phi}$, and $\sigma\ge0$ holds for any closure consistent with the fluctuation--dissipation theorem (Secs.~\ref{sec:hydro} and \ref{sec:modal}), not only for the RTA. Reference~\cite{Zeng2026} also obtains positivity, from the time-correlation structure of its Green--Kubo kernel; the two routes agree on this point and rest on the same property (non-negative spectrum of a stationary correlation), and differ in their starting point, thermodynamic in ours (nonequilibrium chemical potential in phase space, Sec.~\ref{sec:meso}) and statistical-mechanical in theirs. (ii) \emph{Memory.} In the RTA, and in the phonon-gas realization of Ref.~\cite{Zeng2026} [their Eqs.~(7)--(12)], the temporal memory is a sum of exponentials fixed by the eigenvalues of the collision operator; here $\bm{\Phi}$ obeys its own fluctuation--dissipation theorem and may be non-Markovian (sub-/super-Ohmic spectra, Sec.~\ref{sec:spectral}), which is the regime of truncated L\'evy transport in alloys~\cite{Vermeersch2015} and of the low-temperature hydrodynamic window~\cite{Beardo2022}. (iii) \emph{Interface.} Ref.~\cite{Hua2019} treats the interface through an inhomogeneous term fed by the measured spectral transmissivity; Ref.~\cite{Wilson2013} through two channels coupled by fitted rates; Ref.~\cite{Zeng2026} through a finite-rank separable interfacial kernel with illustrative, not material-specific, parameters (their Table~II and Sec.~III.A.2). Here, given a modal boundary condition (thermalizing, or adiabatic below 3~THz; Sec.~\ref{sec:bvp}), the interfacial nonequilibrium resistance follows from the truncation of the long-MFP sector of the same kernel; the two-channel equations are its two-sector projection (Sec.~\ref{sec:twosector}), and $R_{\mathrm{NE}}$ is computed rather than fitted. In this respect the present treatment is closest to Ref.~\cite{Hua2019}, which computes the interfacial response from the BTE with a measured transmissivity. Conversely, the present work does not replace what those treatments supply and which it consumes as input: first-principles lifetimes and the full three-dimensional transducer/substrate geometry with spectral transmissivity~\cite{Hua2019}, Monte Carlo solutions of the complete TDTR geometry~\cite{Ding2014}, the general (non-phonon-gas) Green--Kubo definition of the kernel and its atomistic evaluation for disordered solids~\cite{Zeng2026}, and the transient-grating oscillatory regime analysed in Refs.~\cite{Hua2020,Zeng2026}.

\begin{table*}[t]
\caption{Comparison with previous nonlocal treatments of
phonon heat conduction in the TDTR/TTG regime.}
\label{tab:relation}

\begingroup
\small
\setlength{\tabcolsep}{4pt}
\renewcommand{\arraystretch}{1.18}
\raggedcells

\begin{tabular}{@{}p{\colw}p{\colw}p{\colw}p{\colw}p{\colw}@{}}
\toprule
 & Starting point
 & Kernel / memory
 & Interface
 & Compared with data \\
\midrule

Mahan--Claro~\cite{Mahan1988};\newline
Allen--Perebeinos~\cite{Allen2018}
&
BTE-RTA, linear response
&
$\displaystyle
\sum_\mu
\frac{\kappa_\mu}
{1+(\mathbf{k}\!\cdot\!\bm{\Lambda}_\mu)^2}
$;\newline
exponential
&
--
&
--
\\
\addlinespace

Koh \emph{et al.}~\cite{Koh2014}
&
BTE-RTA, semi-infinite solid, periodic surface heating (1D)
&
same kernel;\newline
exponential
&
prescribed surface $T$ or flux;\newline
no transducer
&
$\Lambda_A(f)$ of Si, Ge, SiGe vs.\ TDTR/BB-FDTR
\\
\addlinespace

Minnich~\cite{Minnich2011};\newline
Ding~\cite{Ding2014}
&
MFP spectroscopy, cutoff model~\cite{Minnich2011};\newline BTE-RTA, MC in 3D TDTR geometry~\cite{Ding2014}
&
suppression function $S(\Lambda/D)$
&
$G$, transmissivity
&
Si $\Lambda_A(D,T)$
\\
\addlinespace

Hua \emph{et al.}~\cite{Hua2014,Hua2019,Hua2020}
&
BTE-RTA Green's function~\cite{Hua2014};\newline generalized Fourier law~\cite{Hua2019};\newline full collision matrix~\cite{Hua2020}
&
same kernel $+$ inhomogeneous term
&
measured\newline
$T_{\mathrm{Si\to Al}}(\omega)$
&
Al/Si TDTR,\newline
no free parameter~\cite{Hua2019}
\\
\addlinespace

Yang--Dames~\cite{YangDames2015};\newline
Xiang \emph{et al.}~\cite{Xiang2022}
&
BTE-RTA (1D analytic);\newline
Guyer--Krumhansl
&
$\Lambda_A(f)$ closed form;\newline
Lorentzian, $\ell$ fitted
&
$G$ fitted
&
TDTR/FDTR of Si and alloys
\\
\addlinespace

Wilson \emph{et al.}~\cite{Wilson2013}
&
two diffusive channels
&
local, coupling $g$
&
$G_1,G_2$ fitted
&
$\Lambda_A(f),G_A(f)$ of SiGe
\\
\addlinespace

Zeng--Dong~\cite{Zeng2026} & Zwanzig/Robertson projection; spatiotemporal Green--Kubo & general; exponential modes in the BTE realization; positivity from correlation structure & finite-rank separable kernel (illustrative) & TTG in Si; Si membrane \\
\addlinespace

This work
&
MNET $+$ Mori--Kubo--Zwanzig\newline
(Liouville)
&
modal Onsager kernel;\newline
any closure obeying the fluctuation--dissipation theorem;\newline
RTA and G--K as limits
&
modal boundary condition $+$ truncation of long-MFP sector;\newline
two-sector projection;\newline
$R_{\mathrm{NE}}$ computed
&
Si $\Lambda_A(D)$, $\Lambda_r$;\newline
Ge/SiGe $\Lambda_A(f)$, $G_A(f)$
\\

\bottomrule
\end{tabular}

\endgroup
\end{table*}
% ======================================================================
\suppsection{Testable predictions}{sec:pred}
(i) \emph{Longitudinal/transverse ratio.} The kernel is isotropic; the anisotropy predicted is that of the \emph{apparent} conductivities under a given probing geometry, not of the material. For an in-plane wavevector $k$ the suppression of the in-plane (longitudinal) and through-plane (transverse) responses is in the ratio $[1-\kappa_L(k)/\kappa_{\mathrm{bulk}}]/[1-\kappa_T(k)/\kappa_{\mathrm{bulk}}]$, which Eqs.~\eqref{eq:FL}--\eqref{eq:FT} bound by 3 in the long-wavelength limit and which the Si spectrum of Fig.~\ref{fig:S1}(a) puts at $1.56$, $1.70$, $1.82$ and $2.01$ for $w_0=1$, $2.4$, $5$ and $12~\mu$m (at the effective wavevectors of the beam-offset width, $k_{\mathrm{eff}}w_0=2.3$--$3.6$, Sec.~\ref{sec:spectra}). The test requires an experiment sensitive to both components at the same in-plane $k$ (beam-offset TDTR with $\Lambda_z$ and $\Lambda_r$ fitted simultaneously~\cite{Feser2012,Wilson2014}, or a variable-spot measurement~\cite{Minnich2011,Ding2014}) in a material whose through-plane response is not dominated by the interface condition, i.e.\ Ge or the alloys at low $f$ rather than Si. In Si, where $\Lambda_z$ is interface-dominated (Sec.~\ref{sec:bvp}), the transverse block appears instead as the difference between the solid and dashed curves of Fig.~\ref{fig:S1}(b), $\approx10$~W\,m$^{-1}$K$^{-1}$ at $D\approx4~\mu$m, growing at smaller $D$. (ii) \emph{Collapse.} $\Lambda_r$ from different experiments~\cite{Minnich2011,Ding2014,Wilson2014} should collapse on the single curve $\kappa_L(k)$ of Fig.~\ref{fig:S1}(a) when each value is plotted against the effective wavevector of its own geometry, obtained from the forward model ($k_{\mathrm{eff}}w_0\simeq2.3$ at $w_0=1~\mu$m, rising to $3.6$ at $10$--$12~\mu$m for the beam-offset width; Sec.~\ref{sec:spectra}); a fixed $c$ would misplace the points by up to $50\%$ in $k$. (iii) \emph{Isotopes.} Isotopically enriched $^{28}$Si gains its extra $\simeq8\%$ of conductivity from modes with $\Lambda\gtrsim3~\mu$m, for which $F_L(2.4\Lambda/w_0)\lesssim10^{-2}$ at $w_0=1~\mu$m: the kernel predicts $\Lambda_r$ unchanged within 1\% between natural and enriched Si while $\kappa_{\mathrm{bulk}}$ differs by 8\%, i.e.\ a larger relative suppression in the purer crystal. (iv) \emph{Temperature.} Through $c_{\mathrm{ph}}(\omega,T)$ and $\tau_\mu(T)$, $\ell_L^2(T)=\tfrac35\langle\Lambda^2(T)\rangle_\kappa$ fixes the temperature dependence of all the above; below $\sim100$~K, where normal scattering dominates, the Markovian closure ceases to apply and the non-Markovian structure of Sec.~\ref{sec:spectral} becomes relevant.

\bibliography{references_v5}